\documentclass[
	reprint,
	superscriptaddress,
	nofootinbib,
	showkeys,
	aps,
	pra,
	longbibliography
]{revtex4-2}

\makeatletter
\def\l@subsubsection#1#2{}
\makeatother

\input{./anc/colors.tex}

\input{./anc/packages.tex}

\allowdisplaybreaks

\usepackage[acronym]{glossaries-extra}
\glssetcategoryattribute{acronym}{nohyperfirst}{true}
\setabbreviationstyle[acronym]{long-short}

\newacronym{rg}{RG}{Renormalization Group}
\newacronym{frg}{FRG}{Functional Renormalization Group}
\newacronym{erg}{ERG}{Exact Renormalization Group}
\newacronym{qft}{QFT}{Quantum Field Theory}
\newacronym{gn}{GN}{Gross--Neveu}
\newacronym{njl}{NJL}{Nambu--Jona--Lasinio}
\newacronym{uv}{UV}{ultraviolet}
\newacronym{ir}{IR}{infrared}
\newacronym{qcd}{QCD}{Quantum Chromodynamics}
\newacronym{twoqcd}{QCD\textsubscript{2}}{two-dimensional Quantum Chromodynamics}
\newacronym{twoym}{YM\textsubscript{2}}{two-dimensional Yang--Mills}
\newacronym{pde}{PDE}{partial differential equation}
\newacronym{ode}{ODE}{ordinary differential equation}
\newacronym{wrt}{w.r.t.}{with respect to}
\newacronym{rhs}{r.h.s.}{right hand side}
\newacronym{lhs}{l.h.s.}{left hand side}
\newacronym{brst}{BRST}{Becchi--Rouet--Stora--Tyutin}
\newacronym{ibp}{i.\,b.\,p.}{integration by parts}

\newcommand{\Reff}{Ref.~}
\newcommand{\Reffs}{Refs.~}

\newcommand{\ie}{\textit{i.e.}}
\newcommand{\eg}{\textit{e.g.}}
\newcommand{\cf}{\textit{c.f.}}

\newcommand{\ii}{\mathrm{i}}
\newcommand{\ee}{\mathrm{e}}

\newcommand{\tr}{\mathrm{tr}}
\newcommand{\Tr}{\mathrm{Tr}}
\newcommand{\Det}{\mathrm{Det}}

\newcommand{\Gk}{\mathit{\Gamma}_k}
\newcommand{\coupling}{g}
\newcommand{\bpsi}{\bar{\psi}}
\newcommand{\bc}{\bar{c}}
\newcommand{\bA}{\bar{A}}

\newcommand{\cpsi}{c_{\psi}}
\newcommand{\tpk}{\Tilde{\partial}_k}

\newcommand{\Nc}{N_{\mathrm{c}}}
\newcommand{\Nf}{N_{\mathrm{f}}}

\newcommand{\dimDirac}{{d_{\gamma}}}
\newcommand{\gammachiral}{\gamma^{\text{ch}}}

\newcommand{\UoneV}{U(1)_\mathrm{V}}
\newcommand{\UoneA}{U(1)_\mathrm{A}}

\newcommand{\vdistance}{\vphantom{\bigg(\bigg)}}
\newcommand{\Vdistance}{\vphantom{\Bigg(\Bigg)}}

\newcommand{\gtilde}{\tilde{g}}
\newcommand{\mtilde}{\tilde{m}}

\DeclareMathOperator\arctanh{arctanh}
\DeclareMathOperator\arccot{arccot}

\newcommand{\cmark}{\ding{51}}
\newcommand{\xmark}{\ding{55}}

\newcommand{\dofun}{\textsc{DoFun}}
\newcommand{\mathematica}{\textsc{Mathematica}}
\newcommand{\formtracer}{\textsc{FormTracer}}

\begin{document}

% % % % % % % % % % % % % % % % % % % % % % % % % % % % % % % % % % % % % % % % % % % % % % % % % %
% % % % % % % % % % % % % % % % % % % % % % % % % % % % % % % % % % % % % % % % % % % % % % % % % %

%\preprint{}

\title{
	Functional renormalization of QCD in \texorpdfstring{$1 + 1$}{1 + 1} dimensions:\texorpdfstring{\\}{ }four-fermion interactions from quark-gluon dynamics
}

%\thanks{comment on the title}

\author{Eric Oevermann \orcidlink{0009-0000-9431-3670}}
	\email{eric.oevermann@uni-jena.de}
	\affiliation{
		Institute for Theoretical Physics, Friedrich Schiller University Jena,
		Max-Wien-Platz 1, 07743 Jena, Germany
	}

\author{Adrian Koenigstein \orcidlink{0000-0001-7482-2195}}
	\email{adrian.koenigstein@uni-jena.de}
	\affiliation{
		Institute for Theoretical Physics, Friedrich Schiller University Jena,
		Max-Wien-Platz 1, 07743 Jena, Germany
	}

\author{Stefan Floerchinger \orcidlink{0000-0002-3428-4625}}
	\email{stefan.floerchinger@uni-jena.de}
	\affiliation{
		Institute for Theoretical Physics, Friedrich Schiller University Jena,
		Max-Wien-Platz 1, 07743 Jena, Germany
	}
	
\date{\today}

\begin{abstract}
	Quantum Chromodynamics in two spacetime dimensions is investigated with the Functional Renormalization Group. 
	We use a functional formulation with covariant gauge fixing and derive Renormalization Group flow equations for the gauge coupling, quark mass and an algebraically complete set of local fermion-fermion interaction vertices. 
	The flow, based on a convenient Callan--Symanzik-type regularization, shows the expected behavior for a super-renormalizable theory in the ultraviolet regime and leads to a strongly coupled regime in the infrared. 
	Through a detailed discussion of symmetry implications, and variations in the gauge group and flavor numbers, the analysis sets the stage for a more detailed investigation of the bound state spectrum in future work. 
\end{abstract}


\maketitle

\tableofcontents

% % % % % % % % % % % % % % % % % % % % % % % % % % % % % % % % % % % % % % % % % % % % % % % % % %
% % % % % % % % % % % % % % % % % % % % % % % % % % % % % % % % % % % % % % % % % % % % % % % % % %

\section{Introduction}

	The study of nonperturbative phenomena in \gls{qft} is challenging but crucial for understanding many intriguing properties of strongly interacting systems in solid state and high-energy physics.
	A prime example is the theory of \gls{qcd} exhibiting confinement, chiral symmetry breaking, hadron formation, multiple thermodynamic phases, \textit{etc.}\ which are all not fully understood yet \cite{Gross:2022hyw}.
	One out of many approaches to make progress on these questions is to investigate lower-dimensional models since they are often more tractable and even admit exact solutions \cite{Schon:2000qy}.
	Usually one can study particular aspects of a theory in this simplified setting and then try to transfer the insights to the more complex case in higher dimensions.

\subsection{Contextualisation}\label{sec:contextualisation}
	
	In this work, we inspect \gls{twoqcd} which is naturally related to \gls{qcd}.
	Here, one can address many of the questions of interest for \gls{qcd} but in a simpler setting. 
	The theory becomes super-renormalizable and possesses much fewer degrees of freedom than in four dimensions, but keeps most of the other important properties of \gls{qcd}, like confinement.
	Yet, it serves as a more realistic toy model for \gls{qcd} than, \eg, the Schwinger model that also shows confinement because of its geometric origin in two dimensions where the Coulomb potential between static (color) charges rises linearly with their distance \cite{Abdalla:1995dm}. 

	\Gls{twoqcd} has been studied in various contexts and for a plenty of reasons.
	In \cref{fig:contextualisation}, we tried to sketch some connections and highlight some motivations.
	For detailed reviews we refer to \Reffs\cite{Abdalla:1995dm,Schon:2000qy,Kalashnikova:2001df}.
	Here, we only list our most important motivations:
		\begin{figure}[h]
			\centering
			\includegraphics[clip, trim = .6cm 22cm 5.7cm .9cm, width = \columnwidth]{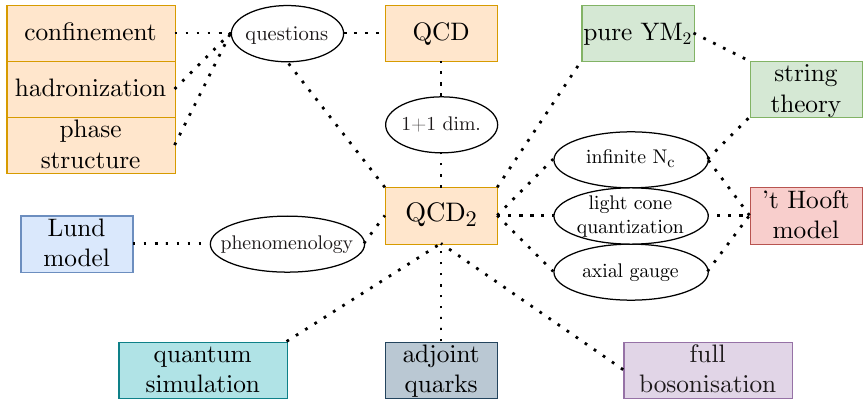}
			\caption[]{
				Overview of various interests in the literature in studying \gls{twoqcd}.
			}
			\label{fig:contextualisation}
		\end{figure}
	Considering just two dimensions is phenomenologically motivated by the Lund (string) model for the dynamics of hadronization \cite{Andersson:1997xwk}. 
	Besides, \gls{twoqcd} has been solved analytically in the limit of an infinite number of colors, ${ \Nc \rightarrow \infty }$, and at one flavor which is referred to as the 't Hooft model \cite{tHooft:1974pnl}. 
	With \gls{qcd}, it shares the properties of confinement, asymptotic freedom, chiral symmetry breaking \cite{Zhitnitsky:1985um} and the existence of a rich bound state spectrum of mesons. 
	Baryons emerge in the next-to-leading order in a ${ 1 / \Nc }$-expansion \cite{tHooft:1973alw} for which \gls{twoqcd} was the first application and test, see also \Reff\cite{Gutierrez:1980gj}. 
	In addition, the 't Hooft model presents an important toy model exploring light-cone quantization \cite{Brodsky:1997de,Mannheim:2020rod}. 
	Much progress has been made on the integrability of the model in this particular limit, ${ \Nc \rightarrow \infty }$. 
	It is still actively investigated, \eg{} in \Reffs\cite{Ambrosino:2023dik,Ambrosino:2024prz,Kochergin:2024quv,Litvinov:2024riz,Hoyer:2025cpf}, and even thermodynamic properties have been explored \cite{Schon:2000qy,Schon:2000vd}. 
	Moreover, this limit establishes a connection to string theory if one concentrates on pure gauge theory \cite{Abdalla:1995dm,Cordes:1994fc,Blau:1993hj}. 
	Interestingly, \gls{twoym} is even solved at finite $ \Nc $ \cite{Migdal:1975zg,Witten:1991we,Witten:1992xu}. 

	We are interested in the model at a finite number of colors and flavors bridging between the 't Hooft model and \gls{qcd}. 
	This case has been studied with (non-Abelian) full bosonization techniques in analogy to approaches to the Schwinger model, see \Reffs\cite{Abdalla:1995dm,Abdalla:1991vua} for reviews and \Reffs\cite{Konik:2021oqv,Lajer:2021kcz} for recent applications. 
	Furthermore, quarks in the adjoint representation of the ${ SU ( \Nc ) }$ gauge group have been considered \cite{Abdalla:1995dm} for which the model behaves very differently in the \gls{ir} becoming deconfining, see, \eg{}, \Reffs\cite{Bergner:2024ttq,Damia:2024kyt} for latest studies. 
	Recently, even quantum simulations of \gls{twoqcd} are subject of discussion \cite{Farrell:2022wyt,Farrell:2022vyh,Liu:2023lsr,Atas:2022dqm,Ciavarella:2023mfc}. 

\subsection{Research objective}

	We take \gls{twoqcd} at finite number of colors and flavors as an exemplary model for investigating the dynamical emergence of bound states.
	Therefore, our medium-term goal is the systematic application and improvement of dynamical partial bosonization techniques using \gls{frg} methods. 
	There are already several \gls{qcd} related works in four dimensions reaching from Fierz-complete truncations to full effective mesonic potentials to which we refer in \cref{sec:four_fermion_interactions}. 
	The four-fermion interactions play a crucial role for the formation of bound-states. 
	In the future, we wish to utilize the simplifications that arise in two dimensions to resolve them in more detail than it is currently feasible in four dimensions. 
	
	This work presents the first \gls{frg} study of \gls{twoqcd} and we establish the very first steps towards this goal. 
	In particular, we demonstrate that simple Callan--Symanzik-type regulators are usable in two dimensions which also motivates our approach since they preserve important symmetries and analyticities. 
	We concentrate on the case of zero temperature and quark chemical potentials but  consider limits of infinite number of colors or flavors too, because a re-derivation of the 't Hooft solution with the \gls{frg} and a connection to other models is a further goal.

	The intermediate achievements of our program that we discuss in this work comprise the study of the minimal ansatz for the effective average action in the gluonic sector and the additional inclusion of a set of Fierz-complete, local four-fermion interactions that are generated from the gauge dynamics.
	This already allows to identify relevant channels for bosonization as well as robust understanding of the dynamics in the \gls{uv} as well as in different limiting cases.

\subsection{Structure}

	This work is structured as follows:
	we first review the microscopic \gls{twoqcd}-action and its symmetries in \cref{sec:model}. 
	Since we deal with a gauge theory, we next survey the steps towards the computation of its quantum effective action in \cref{sec:quantum_effective_action}. 
	This includes the introduction of the background-field method and a discussion of a regularization of Callan--Symanzik type for the fluctuating part of the gauge field. 
	Besides, we briefly explain the \gls{frg} approach and the use of Callan--Symanzik-type regulators in \cref{sec:method}. 
	We are then ready for an initial investigation of the model close to perturbation theory in \cref{sec:minimal_ansatz}. 
	In \cref{sec:four_fermion_interactions}, we include a set of Fierz-complete, local four-fermion interactions in the ansatz for the effective average action. 
	This section can partly be seen as a standalone Fierz-complete study of a purely fermionic theory in two dimensions.
	Bosonization of these interactions is mentioned in the Outlook, \cref{sec:outlook}.
	We conclude in \cref{sec:summary}. 
	More details on certain aspects of the discussion in the main text are provided in the appendix. 

\subsection{Supplementary material}

	Instead of providing all (lengthy) calculations in the main text or adding further detailed appendices, we decided to refer to the (online available) Master's thesis of Eric Oevermann \cite{Oevermann:2024thesis} for technical details and conventions.
	This should make the paper more readable, avoid redundancy, and on the other hand assures that all calculations are available to the reader for the sake of transparency and reproducibility.
	Conventions are the same.

\section{The model}\label{sec:model}

	In this section, we introduce the theory under investigation on a formal level.

\subsection{The microscopic theory}
\label{sec:model_microscopic_action}

	The microscopic action for \gls{twoqcd} in Euclidean spacetime is
		\begin{align}\label{eq:microscopic_action_qcd2}
			S_{\text{QCD\textsubscript{2}}} [ A, \bpsi, \psi ] = \, & \int_{x} \mathcal{L} ( x ) \,	\Vdistance
		\end{align}
	with the Lagrangian
		\begin{align}
			\mathcal{L}
			= \, & \bpsi \, ( \gamma^{\mu} D_\mu + m ) \, \psi + \frac{1}{2\coupling^2} \, \tr ( F_{\mu \nu} F^{\mu \nu} ) \, .	\Vdistance
		\end{align}
	The domain of integration is $\mathbb{R}^2 $ at zero temperature.
	The relation to the action in Minkowski spacetime is discussed in detail, \eg, in \Reff\cite[Sec.~2 and App.~A]{Oevermann:2024thesis}.
	Fermions $\psi$ and anti-fermions $\bpsi$, which we also call quarks and anti-quarks, carry color charge in the fundamental representation of the gauge group and come in $\Nf$ flavors.
	Hence, in index notation they have a color index, ${ c \in \{ 1, \ldots, \Nc \} }$, a flavor index ${ f \in \{ 1, \ldots, \Nf \} }$, and a Dirac index ${ a \in \{ 1, \ldots, \dimDirac \} }$, which we selectively suppress in equations.

	The fermions are supposed to have a bare mass $ m $, which is here approximated to be identical for all flavors. 
	The quarks are coupled to the gauge field $ A $ via the gauge-covariant derivative with matrix-valued gauge fields,
		\begin{align}\label{eq:def_matrix_gauge_field}
			&	D_\mu = \partial_\mu - \ii A_\mu \, ,	&&	A_\mu = A_\mu^z \, T_z \, .	
		\end{align}
	The matrices $ T_z $ with ${ z \in \{ 1, \ldots, \Nc^2 - 1 \} }$ are the hermitian generators of the ${ SU( \Nc ) }$ gauge group (in the fundamental representation) and obey the commutation relation
		\begin{align}
			\big[ T_z, T_w \big] = \ii \tensor{f}{^{v}_{zw}} T_{v} \, ,	\label{eq:su_n_algebra}
		\end{align}
	where $ \tensor{f}{^{v}_{zw}} $ are the structure constants of the Lie algebra. 
	We use the orthogonality relation and normalization
		\begin{align}
			\tr ( T_w \, T_z ) = C(r) \, \updelta_{ w z } \, ,	\label{eq:orthonormality_su_n_generators}
		\end{align}
	where 
	\begin{align}\label{eq:def_C(r)}
		C ( r ) = \, &
		\begin{cases}
			1 / 2 \, ,			&	\text{fundamental representation} \, ,
			\\
			\Nc \, , 					&	\text{adjoint representation} \, .
		\end{cases}
	\end{align}
	In addition, $g$ is the gauge coupling and the components of the field-strength tensor are
		\begin{align}
			F_{\mu \nu} = \partial_\mu A_\nu - \partial_\nu A_\mu - \ii [ A_\mu, A_\nu ] \, .
		\end{align}
	A peculiarity in two dimensions is that the gauge field has no propagating degrees of freedom, which follows from gauge freedom and the equations of motion.
	In the perturbative limit, instead of propagating gluons one has just a Coulomb-like interaction that is mediated by the field $ A_\mu $. 
	In two dimensions, it gives rise to a linear potential between colored charges \cite{Frishman:2010tc,Shifman:2012zz}, see \Reff\cite{Coleman:1985rnk} for a detailed discussion. 
	Already this geometric property leads to an expectation to find color confinement.
	In this sense, we expect signals of color confinement in our results.  
	\newline
	Before we continue, let us briefly comment on the scaling dimensions of the fields in a two-dimensional spacetime.
	In general, fields have energy dimensions
		\begin{align}
			&	[ \psi ] = E^{\frac{d - 1}{2}} \overset{d = 2}{=} E^{\frac{1}{2}} \, ,	&&	[ A ] = E \, .
		\end{align}
	Consequently, one finds 
		\begin{align}\label{eq:energ_dim_gauge_coupling}
			[ g ] = E^{\frac{4 - d}{2}} \overset{d = 2}{=} E
		\end{align}
	for the gauge coupling.
	The fact that the gauge coupling has positive energy dimension makes the theory super-renormalizable in the perturbative domain.

\subsection{Symmetries}

	The model possesses several symmetries in vacuum \cite{Frishman:2010tc}.
	It is constructed to be gauge invariant, where the fermions are supposed to transform under the fundamental representations of the gauge group $SU(\Nc)$. 
	In addition, in Minkowski space, fermions transform as two-dimensional spinors under Lorentz transformations while the gauge field is one form.
	In Euclidean spacetime, the Lorentz group $ SO ( 1, 1 ) $ turns into the rotation group $ SO ( 2 ) $ and the corresponding algebra for the Euclidean gamma matrices is
		\begin{align}
			\big\{ \gamma^\mu, \gamma^\nu \big\} = 2 \, \openone \, \updelta^{\mu \nu}.
		\end{align}
	We again refer to \Reff\cite[Sec.~2 and App.~A~\&~B]{Oevermann:2024thesis} for details on our conventions for the Dirac matrices and on spacetime symmetries in Minkowski and Euclidean space.
	In short, the action is invariant under the continuous proper orthochronous Lorentz transformations as well as the discrete operations of parity, time reversal, and charge conjugation and their respective Euclidean analogues. 
	In addition, the extension to the group of Poincar\'e transformations is a symmetry.

	The fermions also constitute a fundamental representation of the flavor group. 
	If there was no mass term, the action would be invariant under  ${ U_\mathrm{L}( \Nf ) \times U_\mathrm{R}( \Nf ) }$, independent global flavor rotations for the left- and right-handed components. 
	The symmetry reduces to ${ U ( \Nf ) \cong SU ( \Nf ) \times U(1) / \mathbb{Z}_{ \Nf } }$ %[Ex.~8.4]{Wipf2023}
	for non-zero quark masses. 
	Invariance under global phase rotations $ U ( 1 ) $ corresponds to baryon number conservation. 
	Axial transformations $ U_\mathrm{A}( 1 ) $ are no symmetry in presence of the fermion mass, but it is interesting to note that \gls{qcd} is anomaly-free in two dimensions \cite{Shifman:2012zz}, see also \Reff\cite[App.~D]{Oevermann:2024thesis}.
	Hence, a massless one-flavor theory would remain massless (in a perturbative setup) and the presence of chiral symmetry would set important restrictions on possible fermion interactions being generated via quantum fluctuations. 
	
	In any case, the Coleman--Mermin--Wagner--Hohenberg--Berezinskii theorem \cite{Mermin:1966fe,Coleman:1973ci,Hohenberg:1967,Berezinsky:1970fr,Berezinsky:1972rfj} implies that we cannot expect spontaneous symmetry breaking of a continuous global symmetry in our model at finite $ \Nc $ and finite $\Nf$.
	The physical contradiction, if this was possible, would be a growing correlation function of the appearing Goldstone bosons with increasing distances \cite{Shifman:2012zz}.
	There is still the possibility of a Berezinskii--Kosterlitz--Thouless phase-transition \cite{Berezinsky:1970fr,Berezinsky:1972rfj,Kosterlitz:1973xp} at vanishing temperature $ T=0 $ which is not associated with spontaneous symmetry breaking but with a divergence of the correlation length.

	We conclude the symmetry discussion by pointing out that there is no dilatation or conformal symmetry, even for massless quarks, as it can be seen from the positive energy dimension of the gauge coupling \cite{Frishman:2010tc}.
	This is another significant difference to the microscopic theory of \gls{qcd} in ${3 + 1}$ spacetime dimensions.

\section{Quantum effective action}\label{sec:quantum_effective_action}

	We now want to formulate a quantum-field-theoretic description for \gls{twoqcd} by defining its partition function $ Z $ and its \textit{quantum} or \textit{one-particle irreducible effective action} $ \mathit{ \Gamma } $. 
	The main goal of this section is to introduce the known background-field method in some detail such that the advantages of our scheme for regularizing the theory become comprehensible.

\subsection{Gauge fixing, background-field method and the quantum effective action}

	Starting from the microscopic action \labelcref{eq:microscopic_action_qcd2}, the first attempt to define the partition function $ Z $ in presence of sources is to simply integrate over all field configurations,
		\begin{align}
			Z [ \bar{ \eta }, \eta, J ] = \, & \int \mathcal{D} \bpsi \, \mathcal{D} \psi \, \mathcal{D} A \, \exp \Big( - S_{\text{QCD\textsubscript{2}}} [ \bpsi, \psi, A ] \vdistance \notag
			\\
			& + \int_x ( \bpsi \, \eta + \bar{ \eta } \, \psi + J_z^{ \mu } \, A_{ \mu }^z ) \Big) \, . \vdistance
		\end{align}
	We face two challenges:
		\begin{enumerate}
			\item	The path integral takes into account too many field configurations, since some are related by gauge transformations. 
			 		Consequently, $ S_{\text{QCD\textsubscript{2}}}^{ (2) AA } $, the bare ``inverse'' propagator, is in fact not invertible.
			
			\item	Moreover, we must regularize the theory, in particular the \gls{ir}, as we will see.
		\end{enumerate}
	Let us first turn our attention to the first point. 
	We impose a gauge fixing condition,
		\begin{align}
			G [ A ] \stackrel{ ! }{ = } \, & 0 \, 
		\end{align}
	upon the gauge field $ A $ such that in the partition function, 
		\begin{align}
			& Z [ \bar{ \eta }, \eta, J ] = \int \mathcal{D} \bpsi \, \mathcal{D} \psi \, \mathcal{D} A \, \updelta ( G [ A ] ) \, \mathrm{Det} \Big( \frac{ \updelta }{ \updelta \alpha } G [ A ] \Big) \vdistance 
			\\
			& \times \exp \Big( - S_{\text{QCD\textsubscript{2}}} [ \bpsi, \psi, A ] + \int_x ( \bpsi \, \eta + \bar{ \eta } \, \psi + J_z^{ \mu } \, A_{ \mu }^z ) \Big) \, , \vdistance \notag
		\end{align}
	only one representative of each gauge orbit is picked and we dropped the integral over the gauge orbits. 
	The term ${ \updelta G [ A ] / \updelta \alpha }$ denotes the variation \gls{wrt}\ the parameter $ \alpha $ of an infinitesimal gauge transformation. 
	
	In a next step, we want to move on to the quantum effective action $ \mathit{ \Gamma } $. 
	Hereby, we employ the background-field method to construct a manifestly gauge invariant $ \mathit{ \Gamma } $ even though a gauge-fixing procedure has been applied \cite{DeWitt:1965jb,Kallosh:1974yh,Honerkamp:1972fd,Arefeva:1974jv,Sarkar:1974db,Sarkar:1974ni,Kluberg-Stern:1974nmx,Kluberg-Stern:1975ebk,tHooft:1973bhk,Grisaru:1975ei,Abbott:1980hw,Abbott:1981ke}. 
	To begin with, we shift the variable in the path integral
		\begin{align}
			A = \, & \bA + a \, , 
		\end{align}
	where $ \bA $ is a background-field and $ a $, being the fluctuation-field, is the new integration variable,
		\begin{align}
			& Z [ \bar{ \eta }, \eta, J , \bA] = \int \mathcal{D} \bpsi \, \mathcal{D} \psi \, \mathcal{D} a \, \updelta ( G [ \bA, a ] ) \, \vdistance \notag
			\\
			& \quad \times \mathrm{Det} \Big( \frac{ \updelta }{ \updelta \alpha } G [ \bA, a ] \Big) \, \exp \Big( - S_{\text{QCD\textsubscript{2}}} [ \bpsi, \psi, \bA + a ]  \vdistance \notag
			\\
			& \qquad + \int_x [ \bpsi \, \eta + \bar{ \eta } \, \psi + J_z^{ \mu } \, ( \bA_{ \mu }^z + a_{ \mu }^z ) ] \Big) \, . \vdistance
		\end{align}
	Note that the source couples to the full field ${ A = \bA + a }$. 
	In fact, it is equivalent to couple the sources to the background field or to the fluctuation field only, to arrive at the gauge-invariant quantum effective action, see \Reff\cite[Sec.~3.]{Abbott:1980hw}. 

	We now impose the specific gauge fixing condition
		\begin{align}
			0 \stackrel{ ! }{ = } G [ a, \bA ]^z = \, & ( D^\mu[ \bA ] \, a_\mu )^z - \sigma^z \, , 
		\end{align}
	where $ \sigma^z $ is an arbitrary function of the spacetime coordinates. 
	We integrate over all $ \sigma^z $ with a Gaussian weighting around ${ \sigma^z = 0 }$ \cite[Secs.~9.4 \& 16.2]{Peskin:1995ev}, and include the delta-distribution in the action as the term 
		\begin{align}
			& S_{ \mathrm{gf} } [ \bA, a ] = \frac{ 1 }{ \xi } \, \frac{ 1 }{ 2 \coupling^2 } \, \int_x ( D_\mu [ \bA ]\indices{^z_y } \, a^{ \mu y } ) \, ( D_\nu [ \bA ]\indices{_{ z x } } \, a^{ \nu x } )\, . \vdistance
		\end{align}
	The constant $ \xi $ can be chosen arbitrarily. 
	Specifically, the choice ${ \xi = 1 }$ corresponds to the background-field analogue of Feynman--'t Hooft gauge \cite[Sec.~16.6]{Peskin:1995ev}, while ${ \xi = 0 }$ is called the Landau--DeWitt gauge \cite{Dupuis:2020fhh}.

	Furthermore, we express the determinant with the Faddeev--Popov procedure \cite{Faddeev:1967fc}. 
	This leads to the introduction of the ghost fields $ c $ and $ \bar{ c } $ \cite[Secs.~16.2 \& 16.6]{Peskin:1995ev},
		\begin{align}
			& S_{ \mathrm{gh} } [ a, \bA, \bar{ c }, c ] = - \int_x \bar{ c }_z \, D^\mu [ \bA ]\indices{^z_w } \, D_\mu[ a + \bA ]\indices{^w_v } \, c^v \, . \vdistance 
		\end{align}
	This method does not come without problems such as the famous Gribov copies \cite{Gribov:1977wm,Vandersickel:2012tz}, see also \Reff\cite{Dudal:2008xd} for a discussion of Gribov copies in two-dimensional gauge theories. 
	Since we will not access the far \gls{ir} regime of the theory in this paper, these problems are, for the time being, not very relevant to us. 
	The total action now reads 
		\begin{align}\label{eq:gauge_fixed_microscopic_action}
			& S [ \bpsi, \psi, a, \bA, \bar{ c }, c ] \vdistance
			\\
			= \, & S_{\text{QCD\textsubscript{2}}} [ \bpsi, \psi, \bA + a ] + S_{ \mathrm{gf} } [ a, \bA ] + S_{ \mathrm{gh} } [ a + \bA, \bA, \bar{ c }, c ] \, , \notag \vdistance
		\end{align}
	and the partition function is given by
		\begin{align}
			& Z [ \bar{ \eta }, \eta, J , \bA, \bar{ \omega }, \omega ] = \vdistance 
			\\
			= \, & \int \mathcal{D} \bpsi \, \mathcal{D} \psi \, \mathcal{D} a \, \mathcal{D} c \, \mathcal{D} \bar{ c } \, \exp \Big( - S [ \bpsi, \psi, a, \bA, \bar{ c }, c ] + \vdistance \notag
			\\
			& + \int_x [ \bpsi \, \eta + \bar{ \eta } \, \psi + J_z^{ \mu } \, ( \bA_{ \mu }^z + a_{ \mu }^z ) + \bar{ c } \, \omega + \bar{ \omega } \, c ] \Big) \, . \vdistance \notag
		\end{align}
	The first step towards the construction of the quantum effective action is to define the Schwinger functional 
		\begin{align}
			W [ \bar{ \eta }, \eta, J , \bA, \bar{ \omega }, \omega] = \, & \ln Z [ \bar{ \eta }, \eta, J , \bA, \bar{ \omega }, \omega] \, .
		\end{align}
	The quantum effective action $ \mathit{ \Gamma } $ is defined as the Legendre transform of $ W $, 
		\begin{align}
			& \mathit{ \Gamma } [ \bpsi, \psi, a, \bA, \bar{ c }, c ] \vdistance
			\\
			= \, & \sup_{ \eta, \bar{ \eta}, J, \omega, \bar{ \omega } } \, \Big( 
					\int_x [ \bpsi \, \eta + \bar{ \eta } \, \psi + J_z^{ \mu } \, ( \bA_{ \mu }^z + a_{ \mu }^z )  + \bar{ c } \, \omega + \bar{ \omega } \, c ] + \vdistance \notag 
			\\
			& - W [ \bar{ \eta }, \eta, J , \bA, \bar{ \omega }, \omega] 
				\Big) \, . \vdistance \notag
		\end{align}
	We identify the background-field $ \bA $ with the expectation value of the gauge field $ A $, denoted by $ A $ again, after all calculations.
	Not introducing a new variable for this quantum effective action we write
		\begin{align}
			\mathit{ \Gamma } [ \bpsi, \psi, A, \bar{ c }, c ] \equiv \, & \mathit{ \Gamma } [ \bpsi, \psi, a, \bA, \bar{ c }, c ] \vert_{ \bA = A, a = 0 } \, .
		\end{align}

\subsection{Gauge invariance and BRST symmetry}\label{sec:gauge_invariance}

	We now discuss in what sense the background-field method ensures gauge invariance of the quantum effective action. 
	The action $ S_{\text{QCD\textsubscript{2}}} $ is invariant under a gauge transformation of the full field $ A $ which can be distributed in two ways over the fields $ a $ and $ \bA $. 
	There is the \textit{fluctuation-field gauge transformation} acting on the fluctuation-field $ a $ only,
		\begin{align}
			\bar{A}_\mu^z & \mapsto \bar{A}^{\prime z}_\mu = \bar{A}_\mu^z \, ,	\vdistance \label{eq:fluctuation_field_gauge_transformation_background_field}
			\\
			a_\mu^z & \mapsto a^{\prime z}_\mu = a_\mu^z + D_\mu [ a + \bA ]\indices{^z_y} \, \alpha^y \, .	\vdistance \label{eq:fluctuation_field_gauge_transformation_fluctuation_field}
		\end{align}
	Alternatively, in the \textit{background-field gauge transformation} the background-field transforms like a gauge field and the fluctuation-field as a  matter field in the adjoint representation,
		\begin{align}
			\bar{A}_\mu^z & \mapsto \bar{A}^{\prime z}_\mu = \bar{A}_\mu^z + D_\mu [ \bar{A} ]\indices{^z_y} \, \alpha^y \, , \vdistance \label{eq:background_field_gauge_transformation_background_field}
			\\
			a_\mu^z & \mapsto a^{\prime z}_\mu = a_\mu^z + \ii \alpha^y \, ( T_y^{( \mathrm{adj} )} )\indices{^z_w} \, a_\mu^w \, .	\vdistance \label{eq:background_field_gauge_transformation_fluctuation_field}
		\end{align}
	Note that the ghost fields also transform as matter fields in the adjoint representation under background-field transformations. 
	The gauge-fixed action action $ S $ is not invariant under the fluctuation-field gauge transformations \cref{eq:fluctuation_field_gauge_transformation_background_field,eq:fluctuation_field_gauge_transformation_fluctuation_field}. 
	However, it is invariant under the background-field gauge transformations \cref{eq:background_field_gauge_transformation_background_field,eq:background_field_gauge_transformation_fluctuation_field}. 
	As a consequence, the quantum effective action ${ \mathit{ \Gamma } [ \bpsi, \psi, a, \bA, \bar{ c }, c ] }$ is also invariant under the background-field gauge transformations. 

	Finally, the quantum effective action $ \mathit{ \Gamma } [ \bpsi, \psi, A, \bar{ c }, c ] $ inherits gauge invariance under a gauge transformation of the full field $ A $ from the background-field gauge invariance, \cf{} \Reffs\cite{Abbott:1980hw}, \cite[Sec.~5.2.1]{Dupuis:2020fhh}, \cite[Sec.~2]{Freire:2000bq}, \cite[Sec.~4.1]{Gies:2006wv}. 
	It is shown in \Reff\cite[Sec.~3]{Abbott:1980hw} that gauge invariance obtained in this way from the background-field method is equivalent to physical gauge invariance of the full field $ A $ derived without the background-field method. 
	In other words, the equivalence follows from the background-independence of the approach (encoded in Nielsen identities) and Slavnov--Taylor identities \cite[Sec.~5.2.1]{Dupuis:2020fhh}, which we discuss in the following. 

	There is also a residual symmetry of the gauge-fixed action $ S $ in \cref{eq:gauge_fixed_microscopic_action}, the \gls{brst} symmetry \cite{Becchi:1975nq,Tyutin:1975qk}, for which we provide details in \cref{app:BRST_symmetry}.

\subsection{Ward identities}\label{sec:ward_identities}

	We now discuss a few identities which constrain the quantum effective action via the symmetries of the microscopic theory. 
	To start with, before evaluating the quantum effective action ${ \mathit{ \Gamma } [ \bpsi, \psi, a, \bA, \bar{ c }, c ] }$ at ${ \bA = A, \, a = 0 }$, it has to satisfy the \textit{Ward--Takahashi identity} 
		\begin{align}\label{eq:WT_identity}
			\mathcal{ G }^z \mathit{ \Gamma } [ a, \bA ] = \langle \mathcal{ G }^z ( S_{ \mathrm{gf} } + S_{ \mathrm{gh} } ) \rangle_{ J [ \Phi ] } . \vdistance
		\end{align}
	Here, $ \mathcal{ G }^z $ is the generator of the fluctuation field gauge transformations whose explicit form is 
		\begin{align}\label{eq:generator_gauge_transformation}
			\mathcal{ G }^z = \, & D_\mu [ a + \bA ]\indices{^z_w} \, \frac{ \updelta }{ \updelta a_{ \mu w } } 
				+ \ii \, ( T^z \, \psi )^a \, \frac{ \updelta }{ \updelta \psi^a } \vdistance
			\\
			& - \ii \, ( \bpsi \, T^z )_a \, \frac{ \updelta }{ \updelta \bpsi_a }
				+ f\indices{^z_w^v} \, c^w \, \frac{ \updelta }{ \updelta c^v }
				- f\indices{^z_{wv}} \, \bc^w \, \frac{ \updelta }{ \updelta \bc_v }
				\, . \Vdistance	\notag
		\end{align}
	The expression on the \gls{rhs} is evaluated at sources $ J $ that are functionals of the field expectation values. 
	The identity encodes how ${ \mathit{ \Gamma } [ a, \bA ] }$ transforms due to the invariance of the microscopic action $ S_{\text{QCD\textsubscript{2}}} $ under fluctuation-field gauge transformations \cite[Eqs.~(49) and (83), Sec.~4.1]{Gies:2006wv}. 
	These transformations are discussed in \Reff\cite{Freire:2000bq} in more detail.
		
	Similarly, the \gls{brst} symmetry of the gauge-fixed action $ S $ leads to the \textit{Slavnov--Taylor identity} for the quantum effective action \cite[Sec.~5.2.1]{Dupuis:2020fhh}, 
		\begin{align}\label{eq:ST_identity}
			\mathscr{ G } \mathit{ \Gamma } = 0 , \vdistance
		\end{align}
	where $ \mathscr{ G } $ generates the \gls{brst} transformations,
		\begin{align}\label{eq:BRST_generator}
			\mathscr{ G } = \, & ( D_\mu [ a + \bA ] \, c )^z \, \frac{ \updelta }{ \updelta a_\mu^z } 
				+ \ii \, c^z \, ( T_z \, \psi )^a \, \frac{ \updelta }{ \updelta \psi^a } \vdistance  
			\\
			& - \ii \, c^z \, ( \bpsi \, T_z )_a \, \frac{ \updelta }{ \updelta \bpsi_a }
				- \frac{ 1 }{ 2 } \, f\indices{^z_{ u v }} \, c^u \, c^v \, \frac{ \updelta }{ \updelta c^z }
				+ B_z \, \frac{ \updelta }{ \updelta \bc_z }
				\, . \Vdistance	\notag
		\end{align}
	
	The background-independence of the background-field method is encoded in the \textit{Nielsen identity} \cite[Sec.~5.2.1]{Dupuis:2020fhh}, 
		\begin{align}\label{eq:Nielsen_identity}
			\frac{ \updelta \mathit{ \Gamma } [ a, \bA ] }{ \updelta \bA } = 0 , \vdistance
		\end{align}
	which is evaluated on-shell, \ie{} at vanishing sources where $ { \updelta \mathit{ \Gamma } [ a, \bA ] / \updelta a = 0 } $.

\subsection{Regularization via a local regulator}

	We have not yet discussed the regularization of the theory. 
	The need for a \gls{uv}-regularization is not anticipated since the theory is super-renormalizable. 
	However, severe \gls{ir}-divergencies due to massless gauge fields are expected to occur in two dimensions. 

	Even though the following construction points towards the \gls{frg} approach, it only is meant to be a general regularization scheme at this stage. 
	To this end, a regulator piece $ \Delta S_k $ is added to the action $ S $. 
	We specify $ \Delta S_k $ to be of Callan--Symanzik-type for all fields,
		\begin{align}\label{eq:regulator_choice}
			\Delta S_k = \int_x \bigg[ \frac{1}{ 2 \coupling^2 } \, a_\mu^z \, k^2 \, a_z^\mu + \bc_z \, k^2 c^z + c_\psi \, \bpsi \, k \, \psi \bigg] \, .
		\end{align}
	The constant ${ \cpsi \in \mathbb{R} }$ indicates the ratio of regularization scales between quarks and gluons or ghosts. 
	We mostly set it to one in this work.  
	This regulator piece acts as a mass term for all fields and has the following important properties:
		\begin{enumerate}
			\item It is local / momentum independent and invariant under Euclidean rotations. \label{item:locality}
			\item It is invariant under background-field gauge transformations. \label{item:regulator_invariance_gauge_transformations}
			\item It is independent of the background field $ \bA $. \label{item:regulator_invariance_background_field}
		\end{enumerate}
	This leads to the \gls{wrt} the \gls{rg}-scale $ k $ dependent Schwinger functional $ W_k $, 
		\begin{align}
			& W_k [ \bar{ \eta }, \eta, J , \bA, \bar{ \omega }, \omega ] = \ln \bigg[ \int \mathcal{D} \bpsi \, \mathcal{D} \psi \, \mathcal{D} a \, \mathcal{D} c \, \mathcal{D} \bar{ c } \vdistance \notag
			\\
			& \quad \exp \Big( - ( S + \Delta S_k ) [ \bpsi, \psi, a, \bA, \bar{ c }, c ] \vdistance \notag
			\\
			& \quad + \int_x [ \bpsi \, \eta + \bar{ \eta } \, \psi + J_z^{ \mu } \, ( \bA_{ \mu }^z + a_{ \mu }^z ) + \bar{ c } \, \omega + \bar{ \omega } \, c ] \Big) \bigg] \, . \vdistance
		\end{align}
	The so-called \textit{effective average action} $ \Gk $ is obtained by the Legendre transform of $ W_k $ and a subtraction of the regulator term, which is then given in terms of the field expectation values. 
	After all calculations, we identify the background-field $ \bA $ with the expectation value of the gauge field $ A $ as before,
		\begin{align}\label{eq:effective_average_action}
			& \Gk [ \bpsi, \psi, A, \bar{ c }, c ] \vdistance
			\\
			= \, & \bigg[ \sup_{ \eta, \bar{ \eta}, J, \omega, \bar{ \omega } } \Big( 
					\int_x [ \bpsi \, \eta + \bar{ \eta } \, \psi + J_z^{ \mu } \, ( \bA_{ \mu }^z + a_{ \mu }^z )  + \bar{ c } \, \omega + \bar{ \omega } \, c ] \vdistance \notag 
			\\
			& - W_k [ \bar{ \eta }, \eta, J , \bA, \bar{ \omega }, \omega ] 
				\Big)  - \Delta S_k [ \bpsi, \psi, a, \bA, \bar{ c }, c ] 
			\bigg]_{ \bA = A, a = 0 } \, . \vdistance \notag
		\end{align}
	This construction allows to study the theory at arbitrary mass scales $ k $ with the full quantum effective action $ \mathit{ \Gamma } $ being recovered at $ { k = 0 } $. 
	Crucially, the Osterwalder--Schrader axioms \cite{Osterwalder:1973dx,Osterwalder:1974tc} are satisfied by the regularized theory because of property \labelcref{item:locality}. 
	Hence, we can always perform a Wick rotation back to Minkowski spacetime and obtain a well-defined \gls{qft} at all scales $ k $, where in particular Lorentz invariance, causality, and unitarity are fulfilled. 
	Since the regulator piece has no momentum dependences, no additional poles are introduced in the propagators.
	They would complicate the analytic continuation to Minkowski spacetime \cite{Braun:2022mgx} which is connected to a possible breaking of Osterwalder--Schrader reflection positivity \cite{Gurau:2014vwa}. 
	Besides, property \labelcref{item:regulator_invariance_gauge_transformations} importantly ensures that the auxiliary symmetry under background-field gauge transformations remains intact for $ \Gk $. 
	This leads to physical gauge invariance at $ { k = 0 } $ in the way described at the end of \cref{sec:gauge_invariance}. 
	However, one has to adapt how the regulator piece \labelcref{eq:regulator_choice} affects the Ward identities introduced in the previous subsection \labelcref{sec:ward_identities}. 
	Recall that these are linked to fluctuation-field gauge-transformations. 

	It turns out that the Ward--Takahashi identity \labelcref{eq:WT_identity} remains valid at all scales for the mass-like regulator,
		\begin{align}\label{eq:WT_identity_Gk}
			\mathcal{ G }^z \Gk = \langle \mathcal{ G }^z ( S_{ \mathrm{gf} } + S_{ \mathrm{gh} } ) \rangle_{ J } \, . \vdistance
		\end{align}
	This remarkable result is derived in \cref{app:Modified_Ward--Takahashi_identity}. 
	Even though a gluon mass term is not invariant under fluctuation-field gauge transformations, ${ \mathcal{ G }^z \, \Delta S_k }$ happens to be linear in the fluctuation fields only, and the symmetry is restored at all scales by removing the regulator term in \cref{eq:effective_average_action}. 
	Equation \labelcref{eq:WT_identity_Gk} sets constraints on $ \Gk $, \eg{} whether a gluon mass term may (not) appear. 
	In general, the presence of a regulator modifies the identity, \cf{} \cref{eq:modified_WT_identity}. 

	The regulator piece breaks the \gls{brst} symmetry. 
	Since this transformation is highly non-linear, the use of the local regulator \labelcref{eq:regulator_choice} leads to a non-trivial modification of the Slavnov--Taylor identity \labelcref{eq:ST_identity}. 
	See \Reff\cite[Eq.~(78)]{Dupuis:2020fhh} for a general expression of it, which is also called modified Quantum Master Equation. 

	Finally, property \labelcref{item:regulator_invariance_background_field} guarantees that the Nielsen identity \labelcref{eq:Nielsen_identity} holds for the effective average action, too. 
	This statement, 
		\begin{align}\label{eq:modified_Nielsen_identity_CZ}
			\frac{ \updelta \Gk [ a, \bA ] }{ \updelta \bA } \bigg|_{ a = a_{ \mathrm{eq} } } = 0 \, ,
		\end{align}
	again assumes an on-shell evaluation, \ie{} $ a_{ \mathrm{eq} } $ is determined such that ${ ( \updelta \Gk / \updelta a )|_{ a = a_{ \mathrm{eq} } } = 0 }$. 
	At all scales $ k $, $ \Gk $ is on-shell independent of the choice for $ \bA $.
	Consequently, one may identify the background-field with the field expectation value not just at vanishing regulator with $ k = 0 $ but on-shell at arbitrary scales. 
	Off-shell, there is a constraint on $ \Gk $, but it does not contain an explicit scale-dependence via the regulator, \cf{} \cref{eq:on_shell_modified_Nielsen_identity}. 
	The \gls{rhs} of \cref{eq:modified_Nielsen_identity_CZ} is non-zero for arbitrary regulators, \cf{} \cref{eq:modified_Nielsen_identity}.\footnote{
		Note that there seems to be sign mistake in \Reff\cite[Eq.~(79)]{Dupuis:2020fhh}, which is why we re-derive the expression in \cref{app:Modified_Nielsen_identity}. 
		There, we also explain the on-shell evaluation of the resulting expression. 
	}

	To summarize, the local regulator piece \labelcref{eq:regulator_choice} regularizes the \gls{ir}-divergencies of the theory and keeps its analytic structure without violating gauge-invariance and background-field independence in the sense of \cref{eq:WT_identity_Gk,eq:modified_Nielsen_identity_CZ}. 
	These are very special properties of the regulator. 
	Identities related to the latter two would drastically simplify without the need of the gauge-fixing and ghost terms at all in the action.

\section{The method: the FRG}\label{sec:method}

	We need a nonperturbative approach to investigate the infrared properties of the model and in particular its low-energy degrees of freedom. 
	We use the exact, nonperturbative \gls{frg} flow equation \cite{Wetterich:1992yh,Morris:1993qb,Ellwanger:1993mw}, which is also denoted as Wetterich equation or \gls{erg} equation,
		\begin{align}\label{eq:Wetterich_equation}
			& k \, \partial_k  \Gk [ \Phi ] = \frac{1}{2} \, \mathrm{STr}
				\Big(
					( k \, \partial_k R_k )\, \big( \Gk^{(2)} + R_k \big)^{-1} [ \Phi ]
				\Big) \, .	\vdistance
		\end{align}
	It contains a super-trace over all fields of the theory, being collected in an abstract field vector $ \Phi $ and taken to be fixed bare fields, and also accounts for the Grassmann property for fermions. 
	We refer again to \Reff\cite[App.~F]{Oevermann:2024thesis} for our conventions, see also \Reffs\cite{Koenigstein:2023wso,Pawlowski:2005xe,Rennecke:2015lur}.
	For details on the \gls{frg}, we refer to \Reffs\cite{Gies:2006wv,Kopietz:2010zz,Dupuis:2020fhh,Berges:2000ew}. 

	From now on, we emphasize that the regulator function $ R_k $ is supposed to satisfy the usual regulator properties \cite[Eqs.~(13)-(15)]{Gies:2006wv} which is the case for the momentum independent regulator \labelcref{eq:regulator_choice}. 
	For a detailed discussion on choices of regulators, we refer to \Reffs\cite{Litim:2000ci,Litim:2001up,Canet:2002gs,Canet:2003qd,Pawlowski:2005xe,Rosten:2010vm,Osborn:2011kw,Pawlowski:2015mlf,Balog:2019rrg,Braun:2020bhy,Braun:2022mgx,Zorbach:2024zjx}. 

	The main advantages of Callan--Symanzik-type regulators for our study are given by the properties \labelcref{item:locality,item:regulator_invariance_gauge_transformations,item:regulator_invariance_background_field}. 
	Problems with \gls{uv} divergencies might only occur in case of an ansatz for the effective average action $ \Gk $ by means of a truncation that introduces local, dimensionless couplings. 
	We discuss partial bosonization as a solution to this problem with local, four-fermion interaction vertices in the discussion and outlook. 

	The price to be paid with the choice of the fermion regulator is that it obviously breaks chiral symmetry in the limit where quarks are massless.
	It implies that the limit of zero quark mass must be taken very carefully in all expressions.
	One should also not forget that physical interpretations are, strictly speaking, only possible at ${ k = 0 }$ where all symmetries are restored. 
	It is the truncation that we need to apply to the effective action that possibly leads to regulator dependent results.

\section{Quantum effective action: minimal ansatz}\label{sec:minimal_ansatz}

	We must make an ansatz for the effective action for explicit calculations with the Wetterich equation.
	In this section, we stay close to the perturbative setup and solely introduce a scale-dependent gauge coupling, fermion mass, and wave-function renormalization that evolve during the \gls{rg} flow starting off from the classical action.
	Later on, we will significantly extend this truncation.
	We start our discussion with the derivation of the corresponding flow equations.
	Afterwards we discuss their solution and interesting limits.

\subsection{A minimal truncation}

	There are various approximation schemes used in the literature such as expanding the effective action in terms of vertices or field derivatives as well as combinations thereof \cite{Berges:2000ew,Pawlowski:2005xe,Gies:2006wv,Kopietz:2010zz,Dupuis:2020fhh}. 
	The ansatz should always be compatible with the microscopic action such that one can formulate the initial conditions for its scale-dependent quantities.
	In this section, we opt for a truncation to lowest order in fields and derivatives, 
		\begin{align} \label{eq:mirc_eff_action}
			& \Gk [ a, A = \bar{A} + a, \bpsi, \psi, \bc, c ] \vdistance
			\\
			= \, & \int_x 
				\Big( 
					\bpsi \, \big( \gamma^{\mu} \, D_\mu [ A ] + m \big) \, \psi + \frac{1}{4 \coupling^2} \, F_{\mu \nu}^z \, F_z^{\mu \nu}	\vdistance	\notag
					\\ 
					& - \frac{1}{2\xi} \, a_z^\mu \, D_{ \mu w }^z [ \bA ] \, D_{ \nu v }^w [ \bA ] \, a^{ \nu v } - \bc_z \, D_{ \mu w }^z [ \bA ] \, D\indices{^{\mu w}_v} [ A ] \, c^v 
				\Big) \, .	\vdistance	\notag
		\end{align}
	It essentially mimics the microscopic gauge-fixed action, but the gauge coupling $ g $ and the fermion mass $ m $ are scale-dependent and take their bare values in the \gls{uv}.
	Furthermore, we work with renormalized fields that implicitly contain a scale dependence via a wave-function renormalization factor $ Z $ that is initialized at one in the \gls{uv} and absorbed into the fields,
		\begin{align}
			\Phi \equiv \, & ( A, \sqrt{Z_\psi} \, \psi, \sqrt{Z_\psi} \, \bpsi, c, \bc ) \, .
		\end{align}
	As we will see, it is an approximation to not take a ghost wave-function renormalization into account, while it stays an exact statement for the gluon field with a running coupling in our convention as long as gauge invariance holds, see \cref{app:Rescaling of the gauge field}.
	Functional derivatives are always taken \gls{wrt} renormalized fields. 
	The Wetterich equation consequently receives a modification when the renormalized fields are held fixed while taking the scale-derivative, 
		\begin{align}\label{eq:flow_eq_with_an_dim}
			k \, \partial_k  \Gk - \frac{1}{2} \, \eta_{ \mathbf{a} } \, \Phi_{ \mathbf{a} } \frac{\updelta \Gk }{\updelta \Phi_{ \mathbf{a} }}
				= \, & \frac{1}{2} \, \mathrm{STr}
				\bigg( 
					\frac{ k \, \partial_k R_k - \eta \, R_k }{ \Gk^{(2)} + R_k } 
				\bigg) \, . 	\Vdistance
		\end{align}
	Here, we introduced the anomalous dimension
		\begin{align}
			\eta_\psi \equiv - k \, \partial_k  \ln{ Z_\psi} \, , \vdistance	
		\end{align}
	implicitly sum over the field-space vector index $ \mathbf{a} $ and assume that the regulator piece ${ \Delta S_k [ \Phi ] }$ is given in terms of renormalized fields. 
	It is oftentimes useful to again rewrite \cref{eq:flow_eq_with_an_dim} into the form
		\begin{align}\label{eq:flow_eq_log_form}
			k \, \partial_k  \Gk - \frac{1}{2} \, \eta_{ \mathbf{a} } \, \Phi_{ \mathbf{a} } \frac{\updelta \Gk }{\updelta \Phi_{ \mathbf{a} }} 
				= \, & \frac{1}{2} \, \mathrm{STr} 
				\Big( 
					k \, \tpk ( \ln{ \mathscr{P}_k } )
				\Big)	\vdistance
		\end{align}
	by introducing the derivative $\tilde{\partial}_k$, which is supposed to exclusively act on the regulator insertion, as well as the full two-point function 
		\begin{align}
			\mathscr{P}_k [ \Phi ] = \big( \Gk^{(2)} + R_k \big) [ \Phi ] \, .
		\end{align}
	This also allows direct comparison with the perturbative one-loop correction to the effective action and reduces computational complexity in the nonperturbative setup. 
	
	Finally, we also introduce the \gls{rg}-time $ t $ as
		\begin{align}
			&	\partial_t = - k \, \partial_k \, ,	&&	t = - \ln \big( k / \Lambda \big) \in [ 0, \infty ) \, , \vdistance \label{eq:def_rg_time}
		\end{align}
	which is chosen to be positive for the \gls{rg} flow from the \gls{uv} reference scale $ \Lambda $ to the \gls{ir}.

	The third term in the ansatz \labelcref{eq:mirc_eff_action} stems from the gauge fixing and we work in Feynman--'t Hooft gauge ${ \xi = 1 }$ within the present work.
	We plan to investigate the gauge-dependence in future studies.
	After all calculations, one has to evaluate the action at its field expectation values to obtain the effective action. 

	The strategy to derive flow equations for the scale-dependent parameters is as follows: 
	By taking functional derivatives of the Wetterich equation one obtains an expression for the \gls{rg}-time-derivative of an $ n $-point vertex function which can be computed via the \gls{rhs}\ in terms of a loop expression.
	Then one extracts the flow of each scale-dependent parameter via specific projections. 
	In practice, it is useful to do all computations in momentum space.
	Corresponding conventions for Fourier transformations can be found in \Reff\cite[App.~C]{Oevermann:2024thesis}.
	There, we also list all momentum-space expressions for the $ n $-point vertex functions and full propagators that are used during these calculations and that are computed from the ansatz \labelcref{eq:mirc_eff_action}. 

\subsection{Running quark mass and wave function}

	First, we derive equations for $ m $ and $ \eta_\psi $ by taking two functional derivatives of the Wetterich equation \labelcref{eq:flow_eq_with_an_dim} \gls{wrt} the fermionic fields.
	The resulting expression for the fermion two-point vertex function can be depicted in terms of Feynman diagrams as
		\begin{align}\label{eq:flow_ferm_2PT_vertex_diagrams}
			& \partial_t 
			\begin{gathered}
				\includegraphics[clip, trim = .5cm .5cm 3.2cm .5cm]{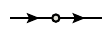}
			\end{gathered}
				\Vdistance
			\\
			= \, &
			\begin{gathered}
				\includegraphics[clip, trim = 0cm 0cm 2.7cm 0cm]{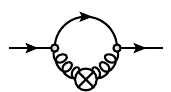}
			\end{gathered}
			+
			\begin{gathered}
				\includegraphics[clip, trim = 0cm 0cm 2.7cm 0cm]{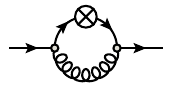}
			\end{gathered} \, ,	\nonumber
		\end{align}
	where the external lines are drawn for visualization purposes only and the regulator insertion is indicated by a crossed circle.
	Gluon propagators are depicted as curly lines and fermion propagators as solid lines with arrows.
	One has to project both sides of \cref{eq:flow_ferm_2PT_vertex_diagrams} on the mass and kinetic term in order to extract flow equations for $ m $ and $ \eta_\psi $, respectively. 
	A detailed derivation is given in \Reff\cite[App.~G]{Oevermann:2024thesis}. 
	
	It turns out that the anomalous dimension vanishes for our choice of gauge fixing and we arrive at
		\begin{align}
			\eta_{\psi} = \, & 0 \, .	\vdistance \label{eq:zero_eta_psi}
		\end{align}
	Besides, it is convenient to introduce dimensionless quantities
		\begin{align}
			&	\gtilde = \frac{g}{k} \, ,	&&	\mtilde = \frac{m}{k} \, . \vdistance \label{eq:def_m_g_tilde}
		\end{align}
	For the sake of better readability, we also provide the following relations,
		\begin{align}
			\partial_t \mtilde = \, & - \partial_k m + \frac{ m }{ k } \, , \qquad \partial_t \gtilde = - \partial_k \coupling + \frac{ \coupling }{ k } \, , \Vdistance \label{eq:derivative_dimless_quantities}
			\\
			k \, \partial_k \bigg( \frac{ k^2 }{ \coupling^2 } \bigg) = \, & 2 \, \frac{ k^2 }{ \coupling^2 } \, \bigg( 1 - \frac{ k }{ \coupling } \, \partial_k \coupling \bigg) = 2 \, \frac{ k^2 }{ \coupling^2 } \, \bigg( \partial_t \ln \gtilde \bigg) \, . \Vdistance \label{eq:gluon_regulator_derivative}
		\end{align} 
	Using this notation we obtain the final result for the flow of the fermion mass for ${ c_\psi = 1 }$,
		\begin{align}\label{eq:flow_mass_full_result} 
			& \partial_t \mtilde - \mtilde + \eta_\psi \, \mtilde \vdistance
			\\
			= \, & \frac{ ( \Nc^2 - 1 ) \, \gtilde^2 }{ 2 \, \uppi \, \Nc \, \mtilde^2 \, ( 2 + \mtilde )^2 } \, \Big( - \mtilde \, ( 2 + \mtilde ) \, \big[ 1 - \eta_\psi \vdistance \notag 
			\\
			& - ( 1 + \mtilde ) \, ( \partial_t \ln \gtilde ) \big] + \big[ -2 \, ( \partial_t \ln \gtilde ) \, ( 1 + \mtilde ) \vdistance \notag 
			\\
			& - ( 2 + \mtilde \, ( 2 + \mtilde ) ) \, ( -1 + \eta_\psi ) \big] \, \ln ( 1 + \mtilde ) \Big) \, . \vdistance \notag
		\end{align}
	Note that the $t$-derivatives of the gauge coupling on the \gls{rhs}\ stem from the dependence of our regulator insertion on the gauge coupling, see \cref{eq:regulator_choice,eq:gluon_regulator_derivative}.
	The prefactor ${ ( \Nc^2 - 1 ) / \Nc }$ is expected because the ${ \Nc^2 - 1 }$ gluons that contribute to the \gls{rhs}\ of \cref{eq:flow_ferm_2PT_vertex_diagrams} must be distributed over the $ \Nc $ color charges carried by the fermions. 

\subsection{Running gauge coupling}

	We use \cref{eq:flow_eq_log_form} to compute the flow of the gauge coupling. 
	This is similar to the perturbative approach, see, \eg, \Reff\cite[Sec.~16.6]{Peskin:1995ev}, and we basically follow these steps applied to the \gls{frg} formalism. 
	The logarithm generates terms of all orders in the background gauge field that are allowed by symmetries.
	The part that is relevant to us is the lowest order background gauge-invariant contribution ${ \tr ( \bar{F}_{\mu \nu} \bar{F}^{\mu \nu} ) }$ which is part of our ansatz. 
	We choose to project onto the part that is quadratic in the background-field $ \bar{A} $. 
	Both sides of the flow equation can then again be represented by Feynman diagrams as 
		\begin{align}\label{eq:flow_gluon_2PT_vertex_diagrams}
			& \partial_t 
			\begin{gathered}
				\includegraphics[clip, trim = .5cm .5cm 2.8cm .5cm]{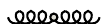}
			\end{gathered}
			= 
			\\
			= \, &	\begin{gathered}
						\includegraphics[clip, trim = .1cm 0cm 2.4cm 0cm]{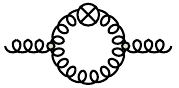}
					\end{gathered}
				- \frac{ 1 }{ 2 }	\begin{gathered}
										\includegraphics[clip, trim = .7cm 0cm 3.1cm .0cm]{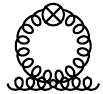}
									\end{gathered}
				+ \notag
			\\
			& -	2 \begin{gathered}
							\includegraphics[clip, trim = 0.2cm 0cm 2.6cm 0cm]{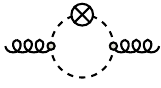}
						\end{gathered}
				+	\begin{gathered}
						\includegraphics[clip, trim = .7cm 0cm 3.1cm .1cm]{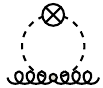}
					\end{gathered}
				+ \notag
			\\
			&	- \Nf \, \begin{gathered}
						\includegraphics[clip, trim = 0.2cm .1cm 2.5cm 0cm]{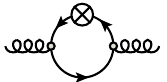}
					\end{gathered}
				+ \frac{ \Nf }{ 2 }	\begin{gathered}
										\includegraphics[clip, trim = .7cm 0cm 3.1cm .1cm]{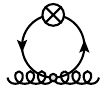}
									\end{gathered}
				\, . \notag
		\end{align}
	Strictly speaking, the quark field has no fundamental vertex that would allow for a ``tadpole''-type diagram. 
	It arises from rewriting
		\begin{align}
			& \Tr \Big( \ln \big[ \gamma^\mu \, D_\mu  + m + k \big] \Big)	\vdistance	
			\\
			= \, &  \frac{1}{2} \ln \Big( \Det \big[ - D^2 + ( m + k )^2 + \ii \, [ D_\mu, D_\nu ] \, \frac{\ii}{4} \, [ \gamma^\alpha, \gamma^\beta ] \big] \Big) \, , \vdistance	\notag
		\end{align}
	where the covariant derivative is \gls{wrt} the background field. 
	For details of this procedure we refer to \Reff\cite[App.~G]{Oevermann:2024thesis}, where furthermore the computation without this rewriting is presented. 
	The advantage is, however, that using Feynman--'t Hooft gauge, ${ \xi = 1 }$, we can treat all three contributions in the same way in terms of a generalized d'Alembert operator,
		\begin{align}\label{eq:generalized_dalembert_operator}
			\mathscr{P}_{ ( s ) }^{ ( r ) } = \, & - \mathbb{ 1 }_{( s )} \,  D^{( r )} [ \bar{A} ]^2 + \bar{F}_{\alpha \beta}^{( r )} \, J_{( s )}^{\alpha \beta} \, .	\vdistance
		\end{align}
	These simplifications are the main advantages of this gauge. 
	The index $ ( s ) $ indicates the spin of the of the field in the field space vector and $ J_{( s )} $ its generator of Lorentz transformations,
	\begin{align} 
		( J_{( 1 )}^{\alpha \beta} )^{\mu \nu} = \, & \ii \left( \updelta^{\alpha \mu} \, \updelta^{\beta \nu} - \updelta^{\alpha \nu} \, \updelta^{\beta \mu} \right) \, , \label{eq:gen_Lorentz_trafo_spin1} \vdistance
		\\
		( J_{( \frac{1}{2} )}^{\alpha \beta} )\indices{^a_b} = \, & \frac{\ii}{4} \, [ \gamma^\alpha, \gamma^\beta ]\indices{^a_b} \, . \label{eq:gen_Lorentz_trafo_spin1half} \vdistance
	\end{align}
	The gauge-group representation of the background-field $ \bar{A} $ is denoted by $ ( r ) $, but we can already identify $ \bar{A} $ with the full field $ A $. 
	The last term in \cref{eq:generalized_dalembert_operator} describes the magnetic-moment interaction with the gauge field. 
	It is the only term from the purely gauge-field ``dynamics'' that contributes to the flow of the gauge coupling in $ 1 + 1 $ dimensions.
	Its overall contribution from the minimal coupling, which is given by the square of the covariant derivative, vanishes. 
	This is because the ghost contribution exactly cancels the gluon contribution, as it should be for the unphysical longitudinal gluon modes and there are no transversal modes in ${ 1 + 1 }$ dimensions. 
	The explicit calculation is provided in \Reff\cite[App.~G]{Oevermann:2024thesis}. 
	We want to highlight that the gauge invariant tensor structure of the form ${ (p^2 \, \updelta^{ \mu \nu } - p^\mu \, p^\nu) }$ is indeed preserved on both sides of the flow \cref{eq:flow_gluon_2PT_vertex_diagrams}. 
	While the ``tadpole''-type diagram does not contribute to the flow of the gauge coupling it is needed to preserve this gauge invariant tensor structure. 
	
	In the end, after adding up all contributions we find
		\begin{align}\label{eq:full_result_running_coupling_main_text}
			\partial_k \coupling = - \partial_t \gtilde + \gtilde = \, & - \gtilde^3 \, \frac{ \frac{ \Nc }{ \uppi } - \frac{ \Nf }{ 12 \, \uppi } \, \frac{ ( 1 - \eta_\psi ) }{ ( 1 + \mtilde )^3 } }{ 1 - \frac{ 11 }{ 12 } \, \frac{ \Nc }{ \uppi } \gtilde^2 } \Vdistance
		\end{align}
	for the flow equation of the gauge coupling.
	Here, we already used ${ c_\psi = 1 }$ and switched to dimensionless quantities \cref{eq:def_m_g_tilde,eq:derivative_dimless_quantities,eq:gluon_regulator_derivative}.
	For general $ c_\psi $, we refer to \Reff\cite[Eq.~(G.37)]{Oevermann:2024thesis}.

	Before we study solutions of the flow equation a few remarks are in order:
	we recover the perturbative one-loop result for the flow of the gauge coupling which is the term at leading order in $ \gtilde $. 
	The negative sign tells that the gauge coupling increases towards lower scales as expected. 
	Besides, we obtain some parts of the higher-loop contributions by expanding the denominator. 
	But of course, there are more terms in the effective action that also contribute at higher order. 
	Examples are terms of higher order in the field-strength tensor or a ghost wave-function renormalization. 
	The pole in the denominator already signals a breakdown of our truncation. 

\subsection{Consistency check: the quark-gluon vertex}\label{sec:consistency_check_quark_gluon_vertex}

	Gauge invariance sets a constraint on the flow of the quark-gluon-vertex. 
	Here, we show that it is indeed satisfied and even in a very convenient way for the particular gauge ${ \xi = 1 }$. 
	The constraint can be checked by taking functional derivatives of the Wetterich equation \gls{wrt} the gluon field and the fermion fields, which diagrammatically reads
		\begin{align}\label{eq:flow_ferm_gluon_3PT_vertex_diagrams}
			& \partial_t 
			\begin{gathered}
				\includegraphics[clip, trim = 2.5cm 0cm 1.cm 0cm]{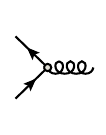}
			\end{gathered}
				\Vdistance
			\\
			= \, &
			\begin{gathered}
				\includegraphics[clip, trim = 2.3cm .1cm 0cm .1cm]{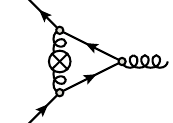}
			\end{gathered}
			+
			\begin{gathered}
				\includegraphics[clip, trim = 2.3cm .1cm 0cm .1cm]{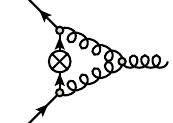}
			\end{gathered} 
			+ \dots \, ,	\nonumber
		\end{align}
	where the dots indicate the other possible regulator insertions. 
	The projection of the \gls{lhs}\ of \cref{eq:flow_ferm_gluon_3PT_vertex_diagrams} onto the quark-gluon-vertex must be proportional to the anomalous dimension of the fermions by construction of the covariant derivative in \cref{eq:def_matrix_gauge_field}, see also \cref{app:Rescaling of the gauge field}. 
	Since it is zero for the gauge ${ \xi = 1 }$, the \gls{rhs}\ must vanish as well which is indeed the case.
	Hence, the constraint is automatically satisfied.

\subsection{Limiting cases}

\begin{figure}
			\subfloat[Finite values ${ \Nc = 3 }$, ${ \Nf = 1 }$.\label{fig:flow_diagrams_finite_values}]{
				\centerline{\includegraphics[clip]{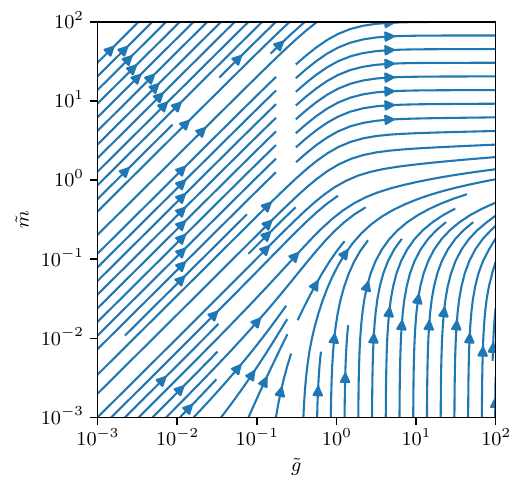}} 
			}
			\newline
			\subfloat[Infinite-$ \Nf $ limit.\label{fig:flow_diagrams_infinite_Nf}]{
				\centerline{\includegraphics[clip]{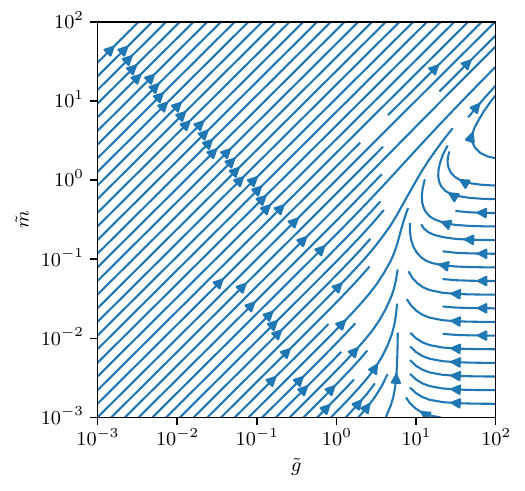}} 
			}
			\caption[Flow of the dimensionless gauge coupling and fermion mass for different flavor numbers.]{
			Flow-diagram for the dimensionless gauge coupling $ \gtilde $ and fermion mass $ \mtilde $ in the $ \gtilde $-$ \mtilde $-plane. 
			To take the infinite-$ \Nf $ limit, we need to re-scale the quantities according to \cref{eq:rescaling_gauge_coupling_large_Nf}. 
			For a given set of values, the arrows indicate the direction of their flow towards the \gls{ir}. 
		}
			\label{fig:flow_diagrams}
		\end{figure}

	In this section, we study the flow equations for the mass and gauge coupling in interesting limits. 
	In agreement with the ansatz we only consider contributions up to order $ \gtilde^2 $. 
	All higher orders are a consequence of the regulator choice for the fluctuating gluon field, see \cref{eq:regulator_choice,eq:gluon_regulator_derivative}. 
	Hence, we take the denominator to be one in \cref{eq:full_result_running_coupling_main_text} and simplify by setting ${ ( \partial_t \ln \gtilde ) \rightarrow 1 }$ in \cref{eq:flow_mass_full_result}.\footnote{In fact, this simply corresponds to an expansion in $g$ up to quadratic order.}
	Furthermore, the anomalous dimension for the fermions $\eta_\psi$ vanishes, \cref{eq:zero_eta_psi}, hence the wave-function renormalization factor is equal to one at all scales, and we work with
		\begin{align}
			& \partial_t \gtilde - \gtilde = - \partial_k \coupling  \Vdistance \label{eq:full_result_running_coupling_leading_order}
			\\
			= \, & \gtilde^3 \, \bigg( \frac{ \Nc }{ \uppi } - \frac{ \Nf }{ 12 \, \uppi } \, \frac{1}{ ( 1 + \mtilde )^3 } \bigg) \, ,	\Vdistance	\notag
			\\
			& \partial_t \mtilde - \mtilde = - \partial_k m \vdistance \label{eq:flow_mass_full_result_leading_order}
			\\
			= \, & \frac{ ( \Nc^2 - 1 ) \, \gtilde^2 }{ 2 \, \uppi \, \Nc \, ( 2 + \mtilde )^2 } \, \frac{2 \, \cpsi \, \ln{\cpsi} }{ \cpsi^2-1 } \, \big[ 2 + \mtilde + \ln ( 1 + \mtilde ) \big] \, . \vdistance \notag
		\end{align}
	These equations are visualized as a flow-diagram in \cref{fig:flow_diagrams}. 
	Generally, such a representation is helpful to identify possible fixed points. 
	We explain the two diagrams and their different regimes further in the subsequent sections.

\subsubsection{Dynamics in the UV}
\label{sec:uv_limit_gauge_coupl_mass}

	Studying the model in the \gls{uv} for large $k$ is relevant because our truncation is expected to be valid in this regime, where the effective average action is supposed to be close to the microscopic action.
	Here, we can analyze how the parameters start to flow due to the fluctuations on largest scales in momentum space.
	The limit is characterized by 
		\begin{align}\label{eq:uv_cond_mtilde_gtilde}
			\mtilde, \gtilde \ll 1 \, . 
		\end{align}
	It is practical to give the limit of \cref{eq:full_result_running_coupling_leading_order,eq:flow_mass_full_result_leading_order} in terms of the variables $ m $, $ \coupling $ and $ k $,
		\begin{align}	
			\partial_k \coupling = \, & - \frac{\coupling^3}{k^3} \, \bar{n} \, , \Vdistance \label{eq:uv_limit_flow_gauge_coupling}
			\\
			\partial_ k m = \, & - \frac{2 \, \cpsi \, \ln{\cpsi} }{ \cpsi^2-1 } \, \frac{ ( \Nc^2 - 1 ) }{ 4 \, \uppi \, \Nc } \, \frac{\coupling^2}{k^2} \, . \Vdistance
		\end{align}
	Here, we (re)introduced the parameter $\cpsi$ and defined
		\begin{align}\label{eq:def_n_bar}
			\bar{n} \equiv \, & \frac{ \Nc }{\uppi} - \frac{\Nf}{ 12 \, \uppi \, \cpsi^2} \, , \Vdistance
		\end{align}
	in order to ease the notation. 
	The flow of the gauge coupling is decoupled from the mass and its solution is
		\begin{align}\label{eq:sol_gauge_coupling_uv}
			\coupling = \, & \frac{ \coupling_\Lambda }{ \sqrt{1 - \bar{ n } \, \coupling_\Lambda^2 \, \big( \frac{1}{k^2} - \frac{1}{\Lambda^2} \big) } } \, . \Vdistance 
		\end{align}
	Thereby, $ \Lambda $ is some \gls{uv} scale and the limit ${ \Lambda \rightarrow \infty }$ can be taken, the theory is super-renormalizable. 
	The gauge coupling approaches a finite \gls{uv} value. 
	That the theory is not free in the \gls{uv} is already clear from the linearly growing potential between quarks which is already present on the classical level as mentioned in \cref{sec:model_microscopic_action}.
	The flow equation for the fermion mass becomes soluble with the result for the gauge coupling.
	We must distinguish according to the sign of $ \bar{n} $:
	\begin{enumerate}
		\item	For the gluon dominated case with ${ \bar{n} > 0 }$ we find
			\begin{align}
				\coupling = \, & \frac{ \coupling_\infty }{ \sqrt{1 - \bar{ n } \, \frac{ \coupling_ \infty^2 }{k^2} } } \, , \Vdistance \label{eq:sol_gauge_coupling_uv_gluon_dominated}
				\\
				m = & \, m_\infty \Vdistance
				\\
				& + \frac{\Nc^2-1}{4\uppi \, \Nc} \, \frac{2 \, \cpsi \, \ln{\cpsi} }{ \cpsi^2-1 } \,  \frac{ \coupling_ \infty }{\sqrt{\bar{n}}} \, \arctanh \Big( \frac{ \sqrt{\bar{n}} \, \coupling_\infty }{ k } \Big) \, , \Vdistance	\notag
			\end{align}
		where the \gls{uv} scale is removed by ${ \Lambda \rightarrow \infty }$, see \Reff\cite[App.~H]{Oevermann:2024thesis} for an explicit derivation.
		The condition ${ \bar{n} > 0 }$ is certainly true for ${ \Nc = 3 }$, ${ \cpsi \geq }1 $ and ${ \Nf \leq 6 }$ -- values similar to \gls{qcd} in four dimensions. 
		It also means that the auxiliary parameter $ \cpsi $ may not take arbitrary values but gluonic and fermionic fluctuations should be introduced at roughly equal scales. 
		As in four dimensions, the gauge coupling increases towards the \gls{ir} and we observe that the gauge coupling develops a singularity, it becomes infinite, when the \gls{rg} scale reaches the ``model scale'',
			\begin{align}
				k =  \sqrt{ \bar{n} } \, \coupling_\infty \, .
			\end{align}
		The truncation breaks down at this scale at the latest and higher-order contributions in $ \gtilde $ become relevant.  
		Taking all higher-order contributions of our simplistic result \cref{eq:full_result_running_coupling_main_text} into account we find that the gauge coupling increases even faster towards lower \gls{rg} scales. 
		It still develops a singularity with infinite derivative but $ \coupling $ remains finite because the denominator in \cref{eq:full_result_running_coupling_main_text} approaches its root fast enough with growing $ \gtilde $. 

		The mass also increases during the flow and develops a singularity at the same scale as the gauge coupling due to the direct dependence on the latter. 
		This is independent from the initial value of the fermion mass. 
		We emphasize that the growing mass in this limit is a feature of the quark regulator that acts as a mass and breaks chiral symmetry. 
		Large values of $ k $ mean that fluctuating quarks are heavy and not that they have large momenta. 
		A perturbative calculation with the diagrams shown in \cref{eq:flow_ferm_2PT_vertex_diagrams} using dimensional regularization does not give a growing mass starting with a massless theory because the projection on the Dirac unit matrix vanishes. 
		Once ${ \mtilde \ll 1 }$ is violated we must use the full equation \cref{eq:flow_mass_full_result}.			
	
		\item	For the fermion dominated case with ${ \bar{n} < 0 }$ we still find
			\begin{align}
				\coupling = \, & \frac{ \coupling_\infty }{ \sqrt{1 - \bar{ n } \, \frac{ \coupling_ \infty^2 }{k^2} } } \, , 
			\end{align}
		while
			\begin{align}
				m = & \, m_\infty + \frac{\Nc^2-1}{4\uppi \, \Nc} \, \frac{2 \, \cpsi \, \ln{\cpsi} }{ \cpsi^2-1 } \, \frac{\coupling_ \infty}{ \sqrt{ | \bar{n} | } }  \Vdistance \notag
				\\
				& \quad \times \Big[ \frac{ \uppi }{ 2 } - \arccot \Big( \frac{ \sqrt{ | \bar{n} | } \, \coupling_ \infty }{ k } \Big) \Big] \, , \Vdistance	
					\end{align}
		where the \gls{uv} scale was again removed ${ \Lambda \rightarrow \infty }$. 
		However, due to the sign-change of $\bar{n}$, instead of running into an \gls{ir} divergence, the coupling decreases towards the \gls{ir} and would be zero at ${ k = 0 }$, if the \gls{uv}-limit was valid on all scales.
		Moreover, the fermion mass increases but approaches a finite \gls{ir} value. 
		In \cref{sec:large_Nc_limit} we continue to discuss this case beyond the approximation of large $k$.	
	\end{enumerate}

	In any case, the UV-regime corresponds to the bottom left corners of the flow-diagrams in \cref{fig:flow_diagrams}. 
	The dimensionless quantities $ \gtilde $ and $ \mtilde $ are considered there and a $ 45^{\circ} $ angle of the curves corresponds to the fact that the ${ 1 / k }$-dependence dominates their flow. 

	We finish the discussion by pointing out two more identities for later use that are valid at large \gls{rg} scales in the \gls{uv}
		\begin{align}
			\eta_\psi \ll \, & 1 \, , \vdistance \label{eq:uv_cond_eta_psi}
			\\
			( \partial_t \ln \gtilde ) = \, & -\frac{1}{ \gtilde } \, \partial_k \gtilde = 1 - \frac{k}{\coupling} \, \partial_k \coupling \stackrel{ \labelcref{eq:full_result_running_coupling_leading_order} }{ \approx } 1 \, . \vdistance \label{eq:uv_cond_gtilde_derivative}
		\end{align}
	The first equation is actually valid on all \gls{rg} scales within our truncation.
	Moreover, it also holds for any truncation in the \gls{uv} regime because it is at least of order $ \gtilde^2 $ and just gives a correction to ${ Z_\psi = 1 }$. 
	Oftentimes, it is even a good approximation at smaller scales or in particular gauges \cite{Jungnickel:1995fp,Gies:2002hq,Gies:2005as}.
	The second identity reflects that the derivative of the fluctuating gluon regulator piece is mainly given by the derivative acting on the $ k^2 $ term and the factor ${ 1/\coupling^2 }$ just gives a correction, see also \cref{eq:gluon_regulator_derivative}. 
	We stress again that this approximation \labelcref{eq:uv_cond_gtilde_derivative} is generally made to simplify the \gls{rhs} of the flow equations within the truncation in this work. 

\subsubsection{Large coupling and fermion mass}
\label{sec:Limit_of_large_coupling_and_fermion_mass}

	Let us return to the case ${ \bar{n} > 0 }$ and ask what actually happens to our full flow equations \labelcref{eq:full_result_running_coupling_leading_order,eq:flow_mass_full_result_leading_order} when the gauge coupling and fermion mass both become large at some intermediate \gls{rg} scale on their evolution to the \gls{ir}.
	An increasing fermion mass in general implies a suppression of fermionic fluctuations which leads to a suppression of their contributions to the \gls{rg} flow. 
	In general, we find that the growth of $ \coupling $ and $ m $ seem to be competing effects in \cref{eq:flow_mass_full_result_leading_order}.

	Thus, assume we reached an regime, where ${ \gtilde, \mtilde \gg 1 }$ and ${ \eta_\psi \ll 1 }$ at some \gls{rg} scale ${ k = \mu }$.
	As a consequence, the flow equations \labelcref{eq:full_result_running_coupling_leading_order,eq:flow_mass_full_result_leading_order} can be approximated by
		\begin{align}
			\partial_k \coupling = \, & - \gtilde^3 \, \frac{ \Nc }{ \uppi } \, , \Vdistance
			\\
			\partial_ k m = \, & - \frac{ \Nc^2 - 1 }{ 2 \, \uppi \, \Nc } \, \frac{2 \, \cpsi \, \ln{\cpsi} }{ \cpsi^2-1 } \, \frac{ \gtilde^2 }{ \mtilde } \, , \vdistance
		\end{align}
	below this scale. 
	Indeed, the fermionic contribution to the flow of the gauge coupling drops out and an analytic solution is available that again allows one to solve the flow equation for the fermion mass. 
	The solutions are
		\begin{align}
			g^2 = \, & \frac{ \coupling_\mu^2 }{ 1 - c \, \big( \frac{ \mu^2 }{ k^2 } - 1 \big) } \, , \Vdistance
			\\
			m^2 = \, & m_\mu^2 - \frac{2 \, \cpsi \, \ln{\cpsi} }{ \cpsi^2-1 } \, \frac{ \Nc^2 - 1 }{ 2 \, \uppi \, \Nc \, ( 1 + c ) } \, \coupling_\mu^2 \Vdistance \nonumber
			\\
			& \times \ln \bigg( \frac{ k^2 }{ \mu^2 } \, \Big[ 1 - c \, \Big( \frac{ \mu^2 }{ k^2 } - 1 \Big) \Big] \bigg)  \, . \Vdistance
		\end{align}
	Here, we introduced 
		\begin{align}
			c \equiv \, & \frac{ \Nc }{ \uppi } \, \frac{ \coupling_\mu^2 }{ \mu^2 }
		\end{align}
	to ease the notation. 
	Both quantities diverge in the limit ${ c \, (\frac{ \mu^2 }{ k^2 } -1 ) \rightarrow 1 }$, meaning that they diverge when $k$ is lowered from $\mu$ to even smaller scales.
	In particular, we find
		\begin{align}
			\lim_{ c \, (\mu^2/k^2 -1 ) \rightarrow 1 } \frac{ m^2 }{ \coupling^2 } = 0 \, ,
		\end{align}
	which means that the coupling grows faster than the fermion mass. 

	The discussion of this section corresponds to the top right corner of the flow-diagram in \cref{fig:flow_diagrams_finite_values}. 
	It appears as if $ \mtilde $ approaches a constant value while $ \gtilde $ increases, but this reflects that $ \gtilde $ increases much faster than $ \mtilde $.

	A plot of the numerical solution to the flow equations \cref{eq:full_result_running_coupling_leading_order,eq:flow_mass_full_result_leading_order} for the boson dominated case is given for the gauge coupling by the blue curve in \cref{fig:flow_g_various_Nf_infty}.

\subsubsection{Infinite-\texorpdfstring{$\Nc$}{Nc} limit}
\label{sec:large_Nc_limit}

	We emphasized already the importance of the sign of $ \bar{n} $ defined in \cref{eq:def_n_bar} in the \gls{uv} regime. 
	Therefore, it is worthwhile to further discuss the model for a boson or fermion dominated theory.
	In this section, we study the former case. 
	In particular the infinite-$ \Nc $ limit, also known as the 't Hooft limit \cite{tHooft:1973alw}, is interesting. 
	The implications for \gls{qcd} are that only so-called ``planar diagrams'' contribute. 
	More general, the limit establishes a connection to a mean-field theory where the fermions do not fluctuate. 
	This statement must be understood in the following sense (quarks still carry a color index and their number grows with $\Nc$): 
	Not all diagrams with fermion propagators drop out of the dynamics.
	Only those diagrams with a purely fermionic loop are suppressed. 
	The limit is taken with ${ \coupling^2 \, \Nc }$ held fixed and $ \Nf $ kept finite.
	Formally, this can be achieved with a rescaling of the gauge coupling 
		\begin{align}\label{eq:rescaling_gauge_coupling_large_Nc}
			\coupling_{ \Nc } \equiv  \sqrt{ \Nc } \, \coupling 
		\end{align}
	before taking the limit with $ \coupling_{ \Nc } $ held fixed. 
	From \cref{eq:full_result_running_coupling_leading_order,eq:flow_mass_full_result_leading_order} we obtain
		\begin{align}
			\partial_k \coupling_{ \Nc } = \, & - \frac{ 1 }{ \uppi } \, \gtilde_{ \Nc }^3 \, ,
		\end{align} 
	for the \gls{rg} flow of the gauge coupling and
		\begin{align}
			\partial_t \mtilde - \mtilde = \, & - \partial_k m \vdistance
			\\
			= \, & \frac{2 \, \cpsi \, \ln{\cpsi} }{ \cpsi^2-1 } \, \frac{ \gtilde_{ \Nc }^2 }{ 2 \, \uppi } \, \frac{ 2 + \mtilde + \ln ( 1 + \mtilde ) }{ ( 2 + \mtilde )^2 } \,  \Vdistance \nonumber
		\end{align}
	for the flow equation of the mass.
	The fermion contribution to the gauge coupling, that has a screening effect, drops out in this limit as expected because the diagrams in \cref{eq:flow_gluon_2PT_vertex_diagrams} with fermion loops are suppressed. This is similar to the scenario discussed previously, where fermion fluctuations were suppressed in the \gls{ir} due to a rising fermion mass. 
	The \gls{uv}-solution \cref{eq:sol_gauge_coupling_uv} becomes exact with ${ \bar{n}_{ \Nc } = 1 / \uppi }$
		\begin{align}
			\coupling_{ \Nc } = \, & \frac{ \coupling_{ \Nc, \Lambda } }{ \sqrt{1 - \bar{ n }_{ \Nc } \, \coupling_{ \Nc, \Lambda }^2 \, \big( \frac{1}{k^2} - \frac{1}{\Lambda^2} \big) } } \, . \Vdistance 
		\end{align}
	The flow equation of the fermion mass does not change qualitatively. 
	The diagrams survive in the limit even though there is a fermion propagator because the external legs are fermionic.

\subsubsection{Large and infinite-\texorpdfstring{$\Nf$}{Nf} limit}

\begin{figure}
			\centering
			\includegraphics[clip, trim = 0cm 0cm 0cm 0cm, width = \columnwidth]{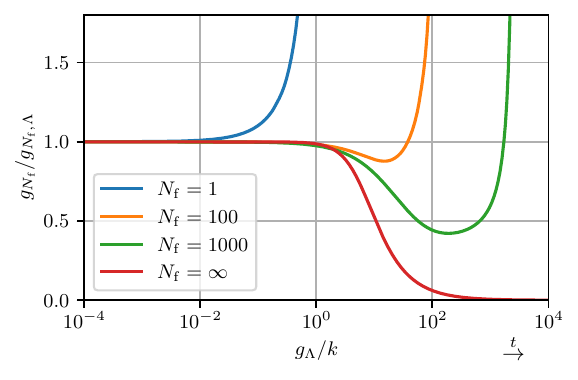}
			\caption[Flow of the gauge coupling $ \coupling_{ \Nf } $ for various flavors $ \Nf $.]{
				Flow of the gauge coupling $ \coupling_{ \Nf } $ for various flavors $ \Nf $. 
				The \gls{uv} values are $ \Lambda / \coupling_{ \Nf, \Lambda } = 10^5 $ and $ m_\Lambda / \coupling_{ \Nf, \Lambda } = 10^{-3} $.
			}
			\label{fig:flow_g_various_Nf_infty}
		\end{figure}

	Now, we turn to the fermion dominated case with ${ \bar{ n } < 0 }$. 
	Similarly, as discussed before, one can take the infinite-$ \Nf $ limit by keeping $ \coupling^2 \, \Nf $ fixed. 
	Again the theory becomes a mean-field theory but this limit is stronger than the infinite-$ \Nc $ limit because gluons do not carry a flavor index. 
	Hence, all diagrams with an internal gluon propagator are suppressed. 
	Formally, this limit is taken by rescaling the gauge coupling 
		\begin{align}\label{eq:rescaling_gauge_coupling_large_Nf}
			\coupling_{ \Nf } \equiv  \sqrt{ \Nf } \, \coupling
		\end{align}
	and after sending ${ \Nf \rightarrow \infty }$ with $ \coupling_{ \Nf } $ held fixed we obtain for \cref{eq:full_result_running_coupling_leading_order,eq:flow_mass_full_result_leading_order}
		\begin{align}
			\partial_k \coupling_{ \Nf } = \, & \gtilde_{ \Nf }^3 \, \frac{ \Nf }{ 12 \, \uppi } \, \frac{ 1 }{ ( 1 + \mtilde )^3 } \, , \vdistance \label{eq:flow_gauge_coupling_large_Nf}
			\\
			\partial_k m = \, & 0 \, . \vdistance \label{eq:solution_mass_infinite_Nf}
		\end{align} 
	As expected, the only diagrammatic contribution that survives is the fermion loop contributing to the flow of the gauge coupling -- here, the bosonic two-point function. 
	The solutions to the infinite-$ \Nf $ flow equations are
		\begin{align}
			\coupling_{ \Nf } = \, & \frac{ \coupling_{ \Nf, \infty } }{ \sqrt{ 1 + \frac{ 1 }{ 12 \, \uppi } \, \frac{ \coupling_{ \Nf, \infty }^2 }{ m_\infty^2 } \, \big( 1 + \frac{ k }{ m_\infty } \big)^{ -2 } } } \, , \Vdistance
			\\
			m = \, & m_\infty \, . \Vdistance
		\end{align} 
	Details on the derivation are given in \Reff\cite[App.~J]{Oevermann:2024thesis}. 
	The mass keeps its constant initial value because the gluon fluctuation in the fermion self energy is suppressed. 
	The gauge coupling decreases and reaches a constant value 
		\begin{align}\label{eq:IR_solution_gauge_coupling_infinite_Nf}
			\coupling_{ \Nf } ( k = 0 ) = \, & \frac{ \coupling_\infty }{ \sqrt{ 1 + \frac{ 1 }{ 12 \, \uppi } \, \frac{ \coupling_{ \Nf, \infty }^2 }{ m_\infty^2 } } } \, , \Vdistance
			\, \vdistance
		\end{align}
	It approaches zero for ${ m_\infty \rightarrow 0 }$. 
	A nonzero mass suppresses the \gls{ir} fluctuations and the coupling remains finite. 
	The solution of this limit is plotted in red in \cref{fig:flow_g_various_Nf_infty}. 
	It is remarkable that the coupling changes significantly over many orders of magnitude and one needs to resolve very low scales.

	If $ \Nf $ is large enough such that the gauge coupling decreases in the \gls{uv} but $ \Nf $ is taken to be finite there is a very different behavior, which is also shown in \cref{fig:flow_g_various_Nf_infty}. 
	The (generated) nonzero mass leads to a suppression of the fermion contribution to the gauge coupling towards the \gls{ir} and the derivative of the coupling changes its sign, see \cref{eq:flow_mass_full_result_leading_order} for ${ \mtilde \rightarrow \infty }$. 
	Therefore, the gauge coupling increases and eventually diverges as discussed in \cref{sec:Limit_of_large_coupling_and_fermion_mass}. 
	The solution given in \cref{eq:solution_mass_infinite_Nf,eq:IR_solution_gauge_coupling_infinite_Nf} acts as a partial fixed point such that the divergence is shifted to lower scales. 
	
	The large-$ \Nf $ limit is depicted in \cref{fig:flow_diagrams_infinite_Nf}. 
	That the arrows take a $ 45^{\circ} $ angle for large values can be explained as follows: even though the dimensionful quantities remain finite, the dimensionless quantities grow as $ k \rightarrow 0 $ and it is again the $ 1 / k $-dependence that dominates as in the UV-regime. 

\subsection{Summary of the chapter}

	To summarize, starting from the finite values of a super-renormalizable theory in the \gls{uv}, the gauge coupling increases due to gluon fluctuations and decreases due to fermionic contributions towards the \gls{ir} while the fermion mass never decreases during the flow. 
	However, coupling and fermion mass always become infinite at a scale ${ k \simeq \coupling_\Lambda }$ at finite $ \Nf $ where the coupling grows much faster than the mass. 
	In the infinite-$ \Nf $ limit, the gauge coupling decreases, the limit $ k \rightarrow 0 $ formally can be taken, and the coupling approaches zero for decreasing initial mass. 

	The results of this section, in particular the increasing gauge coupling in the \gls{ir}, suggest to continue the work in two directions. 
	One step is to improve the truncation in the gauge sector of the effective action to reliably access the regime ${ k \simeq \coupling_\Lambda }$. 
	We need to understand better how what is known about two-dimensional Yang--Mills theory, see \cref{sec:contextualisation}, translates into the formulation we use here. 
	Another possibility is to take higher orders in the fermionic field into account which are driven by the gauge coupling and therefore become very relevant on scales ${ k \simeq \coupling_\Lambda }$. 
	As a first step, we continue with the latter investigation in the next section, which also constitutes a first step in the direction of our overall goal -- obtaining a deeper understanding of dynamical hadronization.

\section{(Local) four-fermion interactions}\label{sec:four_fermion_interactions}

\begin{figure}[h]
			\centering
			\begin{minipage}{.49\linewidth}
				\centering\includegraphics[clip, trim=2.5cm .05cm 0cm 0cm]{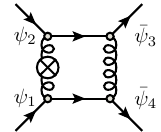}
				%\caption{}
				\label{fig:box_1}
			\end{minipage}
			\hfill
			\begin{minipage}{.49\linewidth}
				\centering\includegraphics[clip, trim=2.5cm .05cm 0cm 0cm]{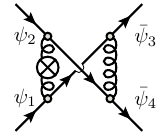}
				%\caption{}
				\label{fig:box_2}
			\end{minipage}
			\hfill
			\begin{minipage}{.49\linewidth}
				\centering\includegraphics[clip, trim=2.5cm .05cm 0cm 0cm]{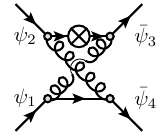}
				%\caption{}
				\label{fig:box_3}
			\end{minipage}
			\hfill
			\begin{minipage}{.49\linewidth}
				\centering\includegraphics[clip, trim=2.5cm .05cm 0cm 0cm]{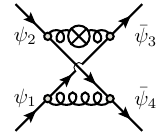}
				%\caption{}
				\label{fig:box_4}
			\end{minipage}
			\caption[Box-diagrams contributing to the flow of the four-fermion couplings.]{
				Box-diagrams contributing to the flow of the four-fermion couplings. 
				One representative regulator insertion is exemplary drawn for each diagram.  
			}
			\label{fig:box_diagrams_four_int}
		\end{figure}

	Quantum fluctuations can dynamically generate effective couplings of all types, as long as these are compatible with the symmetries of the model.
	This comprises also those vertices that are not present in the microscopic action.
	Here, we investigate four-fermion interactions that are directly generated from the fermion-gluon interaction, see \cref{fig:box_diagrams_four_int}.
	From a perturbative point of view, they are the first new interactions that are generated in an expansion in the gauge coupling.
	Based on the previous discussion, they are expected to become relevant at scales ${ k \simeq \coupling_\Lambda }$ in ${1 + 1}$ spacetime dimensions.\footnote{This is different to ${3 + 1}$ dimensional \gls{qcd}, because there the gauge coupling is dimensionless while the four-fermion couplings have nontrivial energy dimension.}
	They are also important to find the emergent \gls{ir} degrees of freedom, because resonant four-fermion channels are associated with the formation of mesonic bound states.
	In this section, we add them to the ansatz of the quantum effective action \labelcref{eq:mirc_eff_action}, derive and solve flow equations for the corresponding couplings. 
	This is the first step to be made to discuss confinement and mesonic bound states.
	We remark that the Schwinger model usually is taken to be the prototype model for two-dimensional gauge theories with the property of a linearly rising Coulomb potential \cite{Abdalla:1995dm}.
	However, the mesonic bound state structure is much richer in \gls{twoqcd} -- at least for ${\Nc \to \infty}$, which makes this model extremely interesting.

	At this point we again emphasize that we are of course not the first to study four-fermion interactions in the context of (models for) \gls{qcd}.
	There are many works on four-fermion theories (in infinite-$ N $ limits) in (${1 + 1}$)-dimensions, which mostly emerged from original works by W.~E.~Thirring \cite{Thirring:1958in}, D.~J.~Gross and A.~Neveu \cite{Gross:1974jv} et al.\ and resulted in a variety of works on bound states, e.g.\ \Reffs\cite{Dashen:1975xh,Lenz:2020cuv}, or the phase structure of these models at nonzero temperature and densities, e.g.\ \Reffs\cite{Dashen:1975xh,Wolff:1985av,Schnetz:2004vr,Basar:2009fg}.
	Also in $3 + 1$ dimensions four-fermion models have been studied extensively to model the strong interaction at low energies.
	Initially, concepts from solid state theory were transferred to particle physics by Y.~Nambu and G.~Jona-Lasinio \cite{Nambu:1961fr,Nambu:1961tp} to study chiral symmetry breaking. 
	Several works have been devoted to Fierz-complete studies of four-fermion interactions in the context of \gls{qcd} in ${3 + 1}$ dimensions in the vacuum \cite{Mitter:2014wpa,Cyrol:2017ewj} as well as at finite temperatures and densities \cite{Braun:2017srn,Braun:2018bik,Braun:2019aow}.
	Other related works in the \gls{frg} framework are \Reffs\cite{Aoki:1999dv,Braun:2009ewx,Braun:2011pp,Braun:2014ata,Aoki:2015hsa,Rennecke:2015eba,Fu:2019hdw,Fu:2022uow,Fu:2024ysj,Fu:2025hcm,Ihssen:2023xlp}.
	However, we are not aware of a study similar to ours, where the emergence of four-fermion interactions is studied in a gauge theory in ${1 + 1}$ dimensions.

\subsection{Extended truncation}

	We improve the minimal ansatz for the quantum effective action \labelcref{eq:mirc_eff_action} from the previous section by including local, momentum-independent fermion interaction terms.
	We restrict ourselves to one flavor, ${ \Nf = 1 }$. 
	The restriction to pointlike interactions without momentum dependence is a severe approximation to describe the \gls{ir} physics as we will discuss later. 
	However, to make progress and in a first step, we add the terms
		\begin{align}
			- & \frac{ 1 }{ \Nc } \int_x \bigg[ \frac{ 1 }{ 2 } \, \Big(
					\lambda_1 \, (\bpsi \, \psi)^2 + \lambda_2 \, (\bpsi \, \gammachiral \, \psi)^2 + \lambda_3 \, (\bpsi \, \gamma^{\mu} \, \psi)^2
				\Big) \notag \vdistance 
			\\
			& + \lambda_4 \, (\bpsi \, T_z \, \psi)^2 + \lambda_5 \, (\bpsi \, \gammachiral \, T_z \, \psi)^2 + \lambda_6 \, (\bpsi \, \gamma^{\mu} \, T_z \, \psi)^2 \notag \vdistance 
			\\
			& + \frac{ 1 }{ 2 } \, \Big( \lambda_7 \, ( \psi^\mathrm{ T } \, \mathcal{C} \, \psi )^{ ( c d ) } \, ( \bpsi \, \mathcal{C} \, \bpsi^\mathrm{ T } )_{ ( d c ) } \notag \vdistance 
			\\
			& + \lambda_8 \, ( \psi^\mathrm{ T } \, \mathcal{C} \, \gammachiral \, \psi )^{ [ c d ] } \, ( \bpsi \, \gammachiral \, \mathcal{C} \, \bpsi^\mathrm{ T } )_{ [ d c ] } \notag \vdistance 
			\\
			& + \lambda_9 \, ( \psi^\mathrm{ T } \, \mathcal{C} \, \gamma^\mu \, \psi )^{ [ c d ] } \, ( \bpsi \, \gamma^\mu \, \mathcal{C} \, \bpsi^\mathrm{ T } )_{ [ d c ] } \Big)	\bigg] \vdistance \label{eq:four-fermion-ansatz}
		\end{align}
	to the ansatz for the effective action in \cref{eq:mirc_eff_action}.
	The matrix $ \gammachiral $ is the two-dimensional analogue to the ($ 3 + 1 $)-dimensional $ \gamma^5 $ and defined in terms of the Euclidean gamma-matrices as
		\begin{align}
			\gammachiral \equiv - \frac{ \ii }{ 2 } \, \epsilon_{ \mu \nu } \, \gamma^\mu \, \gamma^\nu \equiv \, & - \ii \, \gamma^0 \, \gamma^1 \, , \vdistance
		\end{align}
	where $ \epsilon_{ \mu \nu } $ is anti-symmetric with ${ \epsilon^{ 01 } = \epsilon_{ 01 } = 1 }$. 
	We note the following special relation in two dimensions
		\begin{align}
			\bpsi \, \gammachiral \, \gamma^\mu \, \psi = \ii \, \epsilon^{ \mu \nu } \, \bpsi \, \gamma_\nu \, \psi \, . \vdistance
		\end{align}
	Besides, $ \mathcal{C} $ is the charge conjugation operator, see \Reff\cite[App.~A]{Oevermann:2024thesis} for an explicit representation.
	The (anti-)symmetrization of the color indices is denoted by (square) round brackets, \eg
		\begin{align}
			( \psi^\mathrm{ T } \, \mathcal{C} \, \psi )^{ ( c d ) } 
			\equiv \frac{ 1 }{ 2 } \, \Big[
				( \psi^c )^\mathrm{ T } \, \mathcal{C} \, \psi^d + ( \psi^d )^\mathrm{ T } \, \mathcal{C} \, \psi^c
			\Big] \, . \vdistance
		\end{align}
	These bilinear combinations span all four-fermion interaction channels that are compatible with the symmetries and we introduced coupling constants $ \{ \lambda_i \} $ that are dimensionless,
		\begin{align}\label{eq:dimless_four_ferm_couplings}
			[ \lambda_i ] = E^{2 - d} \overset{d = 2}{=} E^0 \, .
		\end{align}
	The bilinears in the first three terms with $ \lambda_1 $, $ \lambda_2 $ and $ \lambda_3 $ are in the color singlet representation of the gauge group, those with $ \lambda_4 $, $ \lambda_5 $ and $ \lambda_6 $ transform in the adjoint representation and the last three terms with $ \lambda_7, \lambda_8 $ and $ \lambda_9 $ correspond to diquark interactions. 
	The diquarks transform in the symmetric or anti-symmetric tensor representation of $ SU( \Nc ) $.
	See \cref{tab:symmetries_quark_bilinears_overview} for an overview of further bilinear symmetry properties and \Reff\cite[App.~B]{Oevermann:2024thesis} for a detailed discussion.
		\begin{table}[h]
			\centering
			\caption{
				Transformation behavior of quark bilinears of mass dimensions ([dim]) $ E^2 $ and $ E^1 $ under Lorentz transformations (L), charge conjugation ($\mathcal{C}$), parity ($\mathcal{P}$), time reversal ($\mathcal{T}$) and $U (1)_{\mathrm{V}/\mathrm{A}}$ phase transformations. 
				($ - $)0 is a (pseudo-)scalar, ($ - $)1 is a (pseudo-)Lorentz vector, $ - $/$ + $ means transformation into itself with negative/positive sign, $ \leftrightarrow ${} is a mapping into partner diquark-bilinear, \cmark (\xmark) means that it is (not) invariant.
			}
			\begin{tabular}{c c c c c c c c c}
				\toprule
				bilinear																		& [dim]						& L						& $ \mathcal{C} $						& $ \mathcal{P} $					& $ \mathcal{T} $					& $ \mathcal{CPT} $							& $ \UoneV $				& $ \UoneA $
				\\
				\midrule
				$ \bpsi \, \gamma^\mu \, \partial_\mu \psi $									& $ E^2 $					& 0						& $ + $									& \phantom{$-$}0					& \phantom{$-$}0					& $ + $										& \cmark					& \cmark
				\medskip \\
				$ \bpsi \, \partial_\mu \psi $													& $ E^2 $					& 1						& $ - $									& \phantom{$-$}1					& \phantom{$-$}1					& $ - $										& \cmark					& \xmark
				\medskip \\
				$ \bpsi \, \gammachiral \, \partial_\mu \psi $									& $ E^2 $					& 1						& $ + $									& $ -1 $							& $ -1 $							& $ - $										& \cmark					& \xmark
				\medskip \\
				$ \bpsi \, \gammachiral \, \gamma^\mu \, \partial_\mu \psi $					& $ E^2 $					& 1						& $ + $									& $ -0 $							& $ -0 $							& $ + $										& \cmark					& \cmark
				\\ \midrule[.2pt]
				$\bpsi \, \gamma^\mu \, \psi $													& $ E^1 $					& 1 					& $ - $									& \phantom{$-$}1 					& \phantom{$-$}1 					& $ + $										& \cmark					& \cmark
				\medskip \\
				$ \bpsi \, \psi $																& $ E^1 $					& 0						& $ + $									& \phantom{$-$}0					& \phantom{$-$}0					& $ + $										& \cmark					& \xmark
				\medskip \\
				$ \bpsi \, \gammachiral \, \psi $												& $ E^1 $					& 0						& $ - $									& $ -0 $							& $ -0 $							& $ - $										& \cmark					& \xmark
				\medskip \\
				$ \bpsi \, \gammachiral \, \gamma^\mu \, \psi $									& $ E^1 $					& 0						& $ - $									& $ -1 $							& $ -1 $							& $ + $										& \cmark					& \cmark
				\\ \midrule[.2pt]
				$ \bpsi \, \gamma^\mu \, \mathcal{C} \, \bpsi^\mathrm{ T } $,					& \multirow{2}{*}{$ E^1 $}	& \multirow{2}{*}{1}	& \multirow{2}{*}{$ - \leftrightarrow$}	& \multirow{2}{*}{ $ -1 $ }			& \multirow{2}{*}{ $ -1 $ }			& \multirow{2}{*}{ $ \leftrightarrow$ }		& \multirow{2}{*}{ \xmark }	& \multirow{2}{*}{ \xmark}
				\\ 
				$ \psi^\mathrm{ T } \, \mathcal{C} \, \gamma^\mu \, \psi $						& & & & & & & &
				\medskip \\
				$ \bpsi \, \mathcal{C} \, \bpsi^\mathrm{ T } $,									& \multirow{2}{*}{$ E^1 $}	& \multirow{2}{*}{0}	& \multirow{2}{*}{$-\leftrightarrow$}	& \multirow{2}{*}{ $ -0 $ }			& \multirow{2}{*}{ $ -0 $ }			& \multirow{2}{*}{ $ - \leftrightarrow$ }	& \multirow{2}{*}{ \xmark }	& \multirow{2}{*}{ \cmark }
				\\ 
				$ \psi^\mathrm{ T } \, \mathcal{C} \, \psi $									& & & &  & & & &
				\medskip \\
				$ \bpsi \, \gammachiral \, \mathcal{C} \, \bpsi^\mathrm{ T } $,					& \multirow{2}{*}{$ E^1 $}	& \multirow{2}{*}{0}	& \multirow{2}{*}{$- \leftrightarrow$}	& \multirow{2}{*}{ \phantom{$-$}0 }	& \multirow{2}{*}{ \phantom{$-$}0 }	& \multirow{2}{*}{ $ -\leftrightarrow$ }& \multirow{2}{*}{ \xmark }	& \multirow{2}{*}{ \cmark }
				\\ 
				$ \psi^\mathrm{ T } \, \mathcal{C} \, \gammachiral \, \psi $					& & & & & & & &
				\medskip \\
				$ \bpsi \, \gammachiral \, \gamma^\mu \, \mathcal{C} \, \bpsi^\mathrm{ T } $,	& \multirow{2}{*}{$ E^1 $}	& \multirow{2}{*}{1}	& \multirow{2}{*}{$- \leftrightarrow$}	& \multirow{2}{*}{ \phantom{$-$}1 }	& \multirow{2}{*}{ \phantom{$-$}1 }	& \multirow{2}{*}{ $ \leftrightarrow$ }	& \multirow{2}{*}{ \xmark }	& \multirow{2}{*}{ \xmark }
				\\ 
				$ \psi^\mathrm{ T } \, \mathcal{C} \, \gamma^\mu \, \gammachiral \, \psi $		& & & & & & & &
				\\
				\bottomrule
			\end{tabular}
			\label{tab:symmetries_quark_bilinears_overview}
		\end{table}

	The nine interaction channels, which one naively writes down to cover all possible four-fermion interactions, form an ``overcomplete basis'' (in the pointlike limit) and the single terms are actually related by Fierz transformations, see \Reff\cite[App.~B]{Oevermann:2024thesis} for details. 
	The minimal set that forms a Fierz-complete basis is given by three channels, for instance those with the bilinears in the singlet or adjoint representation or the diquarks. 
	We remark that if the four-fermion interactions are supposed to be invariant under chiral transformations, this sets additional constraints on the couplings. 
	For instance, choosing the three bilinears in the singlet representation one requires ${ \lambda_1 = - \lambda_2 }$ to obtain the invariant subtraction 
		\begin{align}
			\frac{ 1 }{ 2 \Nc } \, \Big[
				(\bpsi \, \psi)^2 - (\bpsi \, \gammachiral \, \psi)^2
			\Big] \, . \vdistance \label{eq:axial_symmetric_four_ferm_channel}
		\end{align}
	The vector channel $\lambda_3$ is invariant under chiral transformations on its own as can be seen by direct calculation or from the Fierz identities in \Reff\cite[App.~B]{Oevermann:2024thesis}.

\subsection{Connection to other four-fermion models}

	The interaction term \labelcref{eq:axial_symmetric_four_ferm_channel} characterizes what is in four dimensions nowadays called the \gls{njl} model \cite{Nambu:1961tp}.
	It is a very popular model in the context of \gls{qcd} for studying cold deconfined quark matter in an effective low-energy model. 
	It can be generalized to $ \Nf $ flavors with a continuous chiral symmetry \cite{Nambu:1961fr}, see also \Reffs\cite{Braun:2011pp,Andersen:2014xxa,Buballa:2003qv,Springer:2016cji}. 
	The model is named NJL\textsubscript{2} \cite{Thies:2018qgx} or chiral \gls{gn} model in two dimensions. 
	An overview of different models is provided in \Reff\cite{Thies:2020ofv}. 
	D. J. Gross and A. Neveu proposed several models in two dimensions \cite{Gross:1974jv}. 
	They also introduced the term with the coupling $ \lambda_1 $, nowadays referred to as the interaction in the \gls{gn} model in any number of dimensions. 
	It has no mass term on the microscopic level as it is also the case for the NJL\textsubscript{2} model. 
	In two dimensions, the \gls{gn} model is interesting by itself because it shares some properties with (${ 3 + 1 }$)-dimensional \gls{qcd} such as asymptotic freedom, dimensional transmutation and spontaneous (discrete) chiral symmetry breaking \cite[Sec.~I]{Thies:2003kk}. 
	Lastly, the interaction term with $ \lambda_3 $ defines the Thirring model \cite{Thirring:1958in} originally studied for fermions only carrying Dirac indices where all three interaction terms are equivalent and there is an analytical solution available. 
	More remarks on purely fermionic theories are made in \cref{sec:zero_gauge_coupling_four_fermion_int} in the limit of ${ \coupling = 0 }$.

\subsection{Deriving flow equations}

\begin{figure}
			\centering
			\begin{minipage}{.49\linewidth}
				\centering\includegraphics[clip, trim=2.5cm .05cm 0cm 0cm]{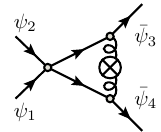}
				%\caption{}
				\label{fig:tri_1}
			\end{minipage}
			\hfill
			\begin{minipage}{.49\linewidth}
				\centering\includegraphics[clip, trim=2.5cm .05cm 0cm 0cm]{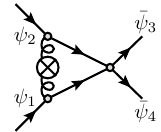}
				%\caption{}
				\label{fig:tri_2}
			\end{minipage}
			\hfill
			\begin{minipage}{.49\linewidth}
				\centering\includegraphics[clip, trim=2.5cm .05cm 0cm 0cm]{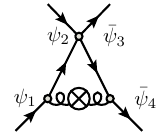}
				%\caption{}
				\label{fig:tri_3}
			\end{minipage}
			\centering
			\begin{minipage}{.49\linewidth}
				\centering\includegraphics[clip, trim=2.5cm .05cm 0cm 0cm]{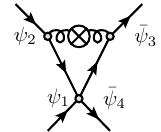}
				%\caption{}
				\label{fig:tri_4}
			\end{minipage}
			\hfill
			\begin{minipage}{.49\linewidth}
				\centering\includegraphics[clip, trim=2.5cm .05cm 0cm 0cm]{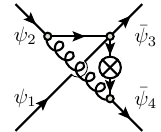}
				%\caption{}
				\label{fig:tri_5}
			\end{minipage}
			\hfill
			\begin{minipage}{.49\linewidth}
				\centering\includegraphics[clip, trim=2.5cm .05cm 0cm 0cm]{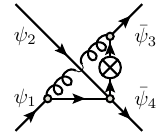}
				%\caption{}
				\label{fig:tri_6}
			\end{minipage}
			\caption[Triangle-diagrams contributing to the flow of the four-fermion couplings. ]{
				Triangle-diagrams contributing to the flow of the four-fermion couplings. 
				One representative regulator insertion is exemplary drawn for each diagram.  
			}
			\label{fig:tri_diagrams_four_int}
		\end{figure}

\begin{figure}
			\centering
			\begin{minipage}{.32\linewidth}
				\centering\includegraphics[clip, trim=2.9cm .05cm .3cm 0cm]{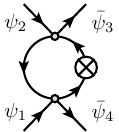}
				\label{fig:ferm_4}
			\end{minipage}
			\hfill
			\begin{minipage}{.32\linewidth}
				\centering\includegraphics[clip, trim=2.5cm .05cm .3cm 0cm]{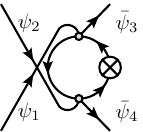}
				\label{fig:ferm_3}
			\end{minipage}
			\hfill
			\begin{minipage}{.32\linewidth}
				$\vcenter{\includegraphics[clip, trim=2.7cm .05cm 0cm 0cm]{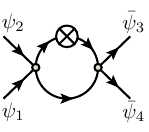}}$
				\label{fig:ferm_1}
			\end{minipage}
			\caption[Purely fermionic diagrams contributing to the flow of the four-fermion couplings.]{
				Purely fermionic diagrams contributing to the flow of the four-fermion couplings. 
				There are just three diagrams in total because regulator insertions in other internal lines lead to the same mathematical expression.  
			}
			\label{fig:ferm_diagrams_four_int}
		\end{figure}

	Next, we study how these four-fermion interactions are dynamically generated by quantum fluctuations in the \gls{frg} formalism.
	To project onto these channels and to generate the corresponding flow equations, we first take functional derivatives of the Wetterich equation \gls{wrt} the fermion fields.
	The \gls{lhs}\ gives the flow of the full four-fermion vertex
	\begin{widetext}
		\begin{align}\label{eq:vertex_four_fermion_channels}
			\frac{ ( 2 \uppi )^8 \, \updelta^4 \, ( \partial_t \Gk ) }{ \updelta \bpsi_{a_4 c_4 f_4} ( - p_4 ) \, \updelta \bpsi_{a_3 c_3 f_3} ( - p_3 ) \, \updelta \psi^{a_2 c_2 f_2} ( p_2 ) \, \updelta \psi^{a_1 c_1 f_1} ( p_1 ) } \Bigg|_{\bar{A}, \, a, \, \psi, \, \bpsi, \, c, \, \bc \, = \, 0} 
			= \, & \partial_t \vcenter{ \hbox{ \includegraphics[clip, trim = 2.5cm 0cm 0cm 0cm]{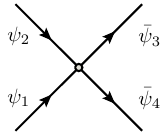} } } \, . \vdistance 
		\end{align}
	\end{widetext}
	Its explicit form with its entire index structure is provided in \Reff\cite[Eq.~C.14]{Oevermann:2024thesis}. 
	The illustration of the \gls{rhs}\ in terms of Feynman diagrams is shown in \cref{fig:box_diagrams_four_int,fig:tri_diagrams_four_int,fig:ferm_diagrams_four_int}.
	In general, we obtain 16 box-type diagrams, 18 triangular-type diagrams and 3 purely fermionic diagrams. 
	The first ones are responsible for the initial generation of the effective couplings. 
	The other diagrams become relevant once the four-fermion couplings are generated and then start feeding back into the flow. 
	
	There are two approaches to extract the flow equations for the couplings from here.

\paragraph{Taking traces with subsequent matrix inversion}

	This approach makes use of computer program assistance.
	This is convenient since the number of diagrams is large and is even growing with every extension of the truncation.
	However, this approach allows to start with an overcomplete basis of interaction channels and to identify the Fierz-complete basis during the procedure.
	Hence, it does not require to work out the correct basis beforehand.

	We proceed as follows.
	First, we make use of the \dofun{} \cite{Huber:2019dkb} software package for \mathematica\ \cite{Mathematica:13.0} in order to take the functional derivatives of the flow equation. 
	Then, we repeatedly contract the open indices with independent projectors using the \formtracer{} software \cite{Cyrol:2016zqb}. 
	One needs nine projectors starting off with the ansatz with nine couplings to end up with a system of nine equations that schematically looks like
		\begin{align}
			M \, ( \partial_t\lambda_1 + 2 \, \eta_\psi \, \lambda_1, \dots, \partial_t\lambda_9 + 2 \, \eta_\psi \, \lambda_9 )^\mathrm{ T } = b \, .
		\end{align}
	Each row is the equation obtained from one projection, $ M $ encodes the pre-factors of the \gls{lhs}\ of the Wetterich equation and the vector $ b $ contains the projected loop expressions from the \gls{rhs}
	To disentangle the \gls{rg}-time derivatives of the nine couplings $ \partial_t \lambda_i $, ${ i \in \{ 1, \ldots, 9 \} }$, on the \gls{lhs}, one has to multiply the system of equations with the inverse of $M$.
	However, this is only possible if $ M $ is invertible which in turn is only possible if the interaction channels form a true basis.
	Hence, noninvertibility of $ M $ directly signals if channels in the ansatz are related by Fierz transformations. 
	The rank of $ M $ determines the number of channels required for a Fierz-complete basis. 
	We find that three are required for the symmetries of our model with ${\Nf = 1}$, and we choose the channels with couplings $ \lambda_1$, $ \lambda_2 $ and $ \lambda_3 $. 
	The full flow equations are given in \Reff\cite[App.~I]{Oevermann:2024thesis}. 

\paragraph{Expansion in tensor structure}

	An alternative approach consists in expanding each loop expression in the tensor structure of the vertex which is only possible if one works with a Fierz-complete basis right from the start. 
	In other words, every product of Dirac or color matrices must be expanded in its basis elements. 
	In the end, one is able to compare both sides of the Wetterich equation and read off the flow equations for the couplings. 
	This procedure is presented for the box-diagrams in \Reff\cite[App.~G]{Oevermann:2024thesis} and was used as a crosscheck for the results obtained via the computer algebra tools.
	This method provides further understanding of the importance to make a Fierz complete ansatz.
	
	We remark that some comparison of the projection and tensor basis approach was discussed and presented in talks already in the context of \Reffs\cite{Braun:2017srn,Braun:2018bik,Braun:2019aow} but not reported on in detail.
	In general, we are not aware of a discussion of the reduction procedure of an ``overcomplete basis'' to a Fierz-complete basis in the context of the \gls{frg} via the Wetterich equation as it is used systematically in the present work.

\subsection{Limiting cases}

	Similar to the previous section, we start our analysis of the \gls{rg} flows with a discussion of particularly interesting limiting cases.
	First, we consider the \gls{uv} limit, where the gauge coupling and fermion mass are assumed to be small compared to the \gls{rg} scale.
	Afterwards, we discuss the full equations to then turn to the limit of vanishing gauge coupling as well as the infinite-$ \Nc $ limit.
	The infinite-$ \Nf $ limit cannot be discussed, because we already restricted ourselves to ${ \Nf = 1 }$.

\subsubsection{Dynamics in the UV}

	First of all, we investigate the \gls{uv} regime as done for the minimal ansatz in \cref{sec:uv_limit_gauge_coupl_mass}. 
	Here, we study how the four-fermion couplings are generated from the dynamics of the gauge sector. 
	In particular, all contributions from the triangle-diagrams and purely fermionic self interactions are irrelevant for the initial condition
		\begin{align}
			\lim_{ k \to \Lambda \to \infty }\lambda_i ( k ) = \, & 0 \, , \quad i = 1,2,3 \, .
		\end{align}
	The contributions from the box-diagrams simplify significantly in the \gls{uv}-limit.
	For details on the calculations we again refer to \Reff\cite[App.~I]{Oevermann:2024thesis}.
	Remember that the latter limit is formally given by \cref{eq:uv_cond_mtilde_gtilde,eq:uv_cond_eta_psi,eq:uv_cond_gtilde_derivative}.
	Ultimately, we find
		\begin{align}
			\partial_t \lambda_1 = \, & \frac{\gtilde^4 \, (\Nc \, [4-(\Nc-2) \, \Nc]+4)}{12 \, \uppi \, \Nc} \, , \Vdistance 
			\\
			\partial_t \lambda_2 = \, & \frac{\gtilde^4 \, [4-\Nc \, (3 \, \Nc+2)]}{12 \, \uppi } \, , \Vdistance 
			\\
			\partial_t \lambda_3 = \, & \frac{\gtilde^4 \, (\Nc+4)}{12 \, \uppi } \, . \Vdistance 
		\end{align}
	All three four-fermion couplings are generated from the gluon-exchange diagrams, \cf{} \cref{fig:box_diagrams_four_int}, but have different flow equations.
	Still, the only differences are the prefactors $ a_i $ and the general form of the flow equations is
		\begin{align}
			\partial_t \lambda_i = \, & a_i \, \gtilde^4 \, , \label{eq:flow_box_schematic_uv}
		\end{align}
	which is expected.
	Axial symmetry of our theory on all scales would require ${ a_1 = -a_2 }$ to ensure that only the symmetry preserving interaction, see \cref{eq:axial_symmetric_four_ferm_channel}, is generated in addition to the coupling $ \lambda_3 $.
	As can be seen from the prefactors, this is not the case already in the \gls{uv}, which implies that chiral symmetry -- despite being a symmetry of the path integral -- seems to be broken by the \gls{rg} flow equations.
	In fact, we trace this symmetry breaking back to the choice of the fermion regulator in \cref{eq:regulator_choice}, which breaks chiral symmetry explicitly.
	We come back to this regulator feature below.

	However, using the \gls{uv}-solution to the gauge coupling, \cref{eq:sol_gauge_coupling_uv}, we can analytically integrate \cref{eq:flow_box_schematic_uv} and find 
		\begin{align}
			\lambda_i = \, & -\frac{ a_i }{ 2 \, \bar{n}^2 } \, \bigg[ \bigg( 1 - \frac{ k^2 }{ \bar{n} \, \coupling_\infty^2 } \bigg)^{-1} - \ln \bigg( 1 - \frac{ \bar{n} \, \coupling_\infty^2 }{ k^2 } \bigg) \bigg] \Vdistance \nonumber
			\\
			= \, & \frac{ a_i }{ 4 } \, \frac{ \coupling_\infty^4 }{ k^4 } + \mathcal{O} \bigg( \frac{ \coupling_\infty^6 }{ k^6 } \bigg)  \, , \Vdistance
		\end{align}
	where the limit ${ \Lambda \rightarrow \infty }$ is implied. 
	The couplings are driven by the gauge coupling and start to grow significantly once the scale $ k $ reaches the scale of the gauge coupling in the \gls{uv}, $ \coupling_\infty $. 
	If this solution was valid on all scales, the couplings would become infinite in the limit ${ k^2 / ( \bar{n} \, \coupling_\infty^2 ) \rightarrow 1 }$ where the gauge coupling diverges. 
	However, the equations are of course no longer valid in this regime.

\subsubsection{Full equations}
\label{sec:numerical_study_full_equations_four_ferm_couplings}

\begin{figure}
		\centering
		\begin{minipage}{0.8\linewidth}
			\raggedleft
			\includegraphics[clip, scale=.909]{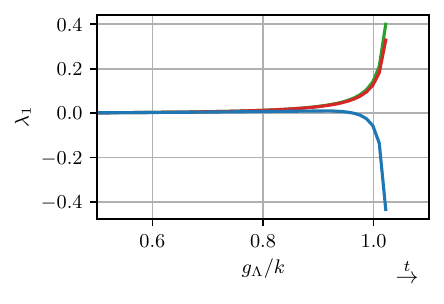}\\
			\includegraphics[clip, scale=.909]{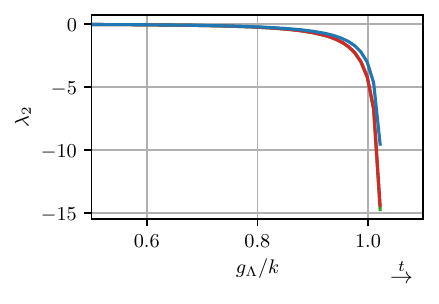}\\
			\includegraphics[clip, scale=.909]{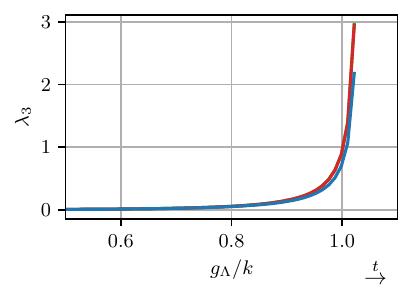}
		\end{minipage}
			\caption[Flow of the four-fermion couplings separated into the different diagrammatic contributions. ]{
					Numerical solution to the flow equations of the four-fermion couplings is depicted with $ m_\Lambda / \coupling_\Lambda = 10^{ -2 } $ and $ \lambda_i = 0 $ at scale $ \Lambda / \coupling_\Lambda = 10^5 $ and $ \Nc = 3 $. 
					Contributions from the box-diagrams in \cref{fig:box_diagrams_four_int} only are shown in blue, adding those from the triangle-diagrams from \cref{fig:tri_diagrams_four_int} in red and again adding the purely fermionic diagrams from \cref{fig:ferm_diagrams_four_int} in green.
					}
			\label{fig:flow_four_ferm_coupling}
		\end{figure}

	To obtain the full solution to our system of flow equations, see again \Reff\cite[App.~I]{Oevermann:2024thesis}, including the flow equations of the gauge coupling and fermion mass, \cref{eq:full_result_running_coupling_leading_order,eq:flow_mass_full_result_leading_order}, we have to make use of numerical integration. 
	The \gls{uv}-limit helps to test the numerical implementation against an analytical solution. 
	The numerical results are shown in \cref{fig:flow_four_ferm_coupling} where we distinguish between the contribution from all diagrams, the box-diagrams plus triangle-diagrams, and the box-diagrams only. 
	We find that all couplings are driven to $ \pm \infty $ by the gauge coupling that develops its singularity.
	The box-diagrams (two gluon exchange diagrams) are the most important in the \gls{uv} regime because they generate the couplings and dominate the flows as long as their values are small compared to $ \gtilde $.
	The triangle-diagrams become more relevant at scales ${ k / \coupling_\Lambda \simeq 1 }$ and lead to a stronger growth of the couplings.
	Their impact is crucial for the coupling $ \lambda_1 $. 
	There is a sign change in the flow which seems to be a special feature for ${ \Nc = 3 }$. 
	Knowledge about the sign is important to the flow equations of the purely fermionic diagrams.
	At this point, it is not clear yet how much they are of importance during the flow. 
	Their contribution is barely visible, but integrating towards even lower scales becomes problematic where we expect our truncation to break down.

\subsubsection{Limit of zero gauge coupling}\label{sec:zero_gauge_coupling_four_fermion_int}

	Let us also consider the limit of vanishing gauge coupling, ${ \coupling \rightarrow 0 }$, which reduces the model to a purely four-fermion theory in ${1 + 1}$ dimensions.
	This is of interest for several (related) reasons: 
		\begin{enumerate}
			\item The purely fermionic model alone is of interest as a standalone \gls{qft}. 
				It treats a Fierz-complete set of four-fermion interactions and one can learn, amongst other things, about the validity of a one or few channel approximation like the \gls{gn} model. 			
			\item The divergence in the gauge coupling prevents us from gaining insights in the dynamics caused by the purely fermionic diagrams shown in \cref{fig:ferm_diagrams_four_int}. 
			The simplest way to study these contributions is to manually ``switch off'' the gauge sector by setting ${ \coupling = 0 }$.
			Another approach, which we do not follow within this work, would be to artificially regularize the gauge sector in a way that renders the flow of the gauge coupling finite and nonzero in the \gls{ir}.
			\item The limit corresponds to a scenario where the purely fermionic diagrams are the dominant ones. 
				This may indeed be the case for a more sophisticated truncation of the effective action in its gauge sector that renders the flow of the dimensionless gauge coupling finite. 
				In four dimensions, this has been achieved \cite{Gies:2002af} and we expect a similar behaviour for two dimensions.
		\end{enumerate}
	Hence, the flow equations that we consider in the following are,
		\begin{align}
			& \partial_t \lambda_1 = \frac{ (\lambda_1)^2 \, \Nc - ( \lambda_1 + \lambda_2 ) \, ( \lambda_1 + 2 \, \lambda_3 ) }{ \uppi \, \Nc \, ( 1 + \mtilde ) } \, , \Vdistance \label{eq:flow_zero_gauge_coupling_one}
			\\
			& \partial_t \lambda_2 = - \frac{ (\lambda_2)^2 \, \Nc - (\lambda_1 + \lambda_2) \, (\lambda_2 - 2 \, \lambda_3) }{ \uppi \, \Nc \, ( 1 + \mtilde ) } \, , \Vdistance \label{eq:flow_zero_gauge_coupling_two}
			\\
			& \partial_t \lambda_3 = -\frac{ \lambda_1 \, \lambda_2 }{ \uppi \, \Nc  \, ( 1 + \mtilde ) } \, . \Vdistance \label{eq:flow_zero_gauge_coupling_three}
		\end{align}
	For a derivation, see \Reff\cite[App.~I]{Oevermann:2024thesis}.
	Here, we assume a vanishing anomalous dimension which we justify in \cref{sec:contribution_to_fermion_two_pt_vertex}. 
	Furthermore, we do not include the flow of the fermion mass. 
	Its flow is in fact not \gls{uv}-finite which we elaborate on in \cref{sec:contribution_to_fermion_two_pt_vertex}.
	Anyways, the mass is not very important in the present discussion because it only appears in a global pre-factor and can be absorbed by a redefinition of the \gls{rg} time and \gls{rg} scale.

	Inspecting \cref{eq:flow_zero_gauge_coupling_one,eq:flow_zero_gauge_coupling_two,eq:flow_zero_gauge_coupling_three}, we recognize the flow equation of the two-dimensional \gls{gn} model in the limit ${ \lambda_2 = \lambda_3 = 0 }$.
	We also find that the flow equation of $ \lambda_3 $ is a bit special because there is no $ (\lambda_3)^2 $ term. 
	We presume that this is a specific result of our particular regularization scheme in ${ d = 2 }$ spacetime dimensions. 
	This is pointed out below in \cref{sec:regulator_artefacts} in more detail.

	The key feature in two dimensions is, that there is no term linear in the couplings in the flow equations that would stem from rescaling the couplings to dimensionless quantities. 
	As expected from the Fierz complete ansatz, there are terms that couple the equations nonlinearly. 
	However, if there is only one of the couplings that is nonzero, it does not generate the other couplings and working with one channel is exact. 
	Another remarkable observation is that if ${ \lambda_1 = -\lambda_2 }$ is satisfied at some scale, it holds at all scales. 
	Hence, even though the regulator insertion does not respect chiral symmetry, it can be restored in the flow of the four-fermion couplings, see also \cref{eq:axial_symmetric_four_ferm_channel}.
	We conclude that, in the full setup, the gluon exchange diagrams are actually responsible for the breaking of the chiral symmetry with the fermionic regulator.
	
\begin{figure}
			\centering
			\begin{minipage}{0.8\linewidth}
				\raggedleft
				\includegraphics[clip, scale=.909]{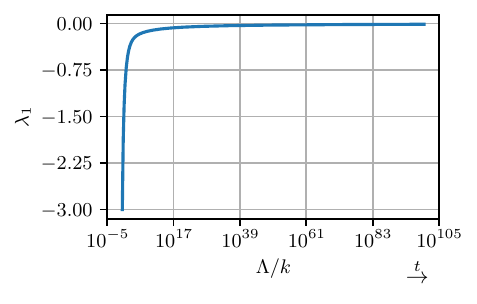}\\
				\includegraphics[clip, scale=.909]{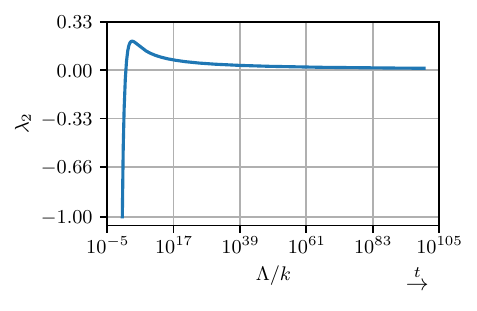}\\
				\includegraphics[clip, scale=.909]{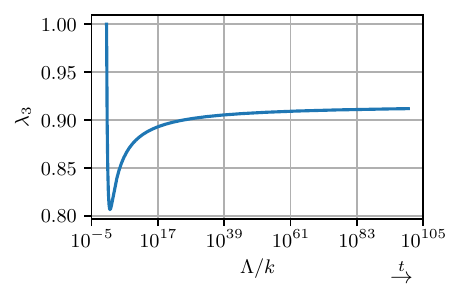}
			\end{minipage}
			\caption[Fixed point flow of the four-fermion couplings in purely fermionic dynamics.]{
					Flow of the four-fermion couplings into a fixed point in purely fermionic dynamics. 
					At ${ t = 0 }$ the values are ${ \lambda_i = ( -3.0, -1.0, 1.0 ) }$.
						}
				\label{fig:fixed_point_fermionic_flow_four_ferm_coupling}
		\end{figure}

	The couplings must be initialized at nonzero values to obtain a nontrivial flow. 
	The case ${ \lambda_1 = \lambda_2 = \lambda_3 = 0 }$ corresponds to a Gaussian fixed point, if there are no gluons present. 
	The class of values where ${ \lambda_1 = \lambda_2 = 0 }$ contains the only fixed points of the model. 
	It has an attractive direction which is exemplary shown in \cref{fig:fixed_point_fermionic_flow_four_ferm_coupling} and derived in greater detail in \Reff\cite[Sec.~5.3.3]{Oevermann:2024thesis}. 
	However, starting from initial conditions, where the couplings approach a fixed point in the \gls{ir}, one can show that the couplings diverge for an integration towards the \gls{uv}. 
	This behavior is known in the \gls{gn} model as well. 
	In contrast, the mutual coupling of the flow equations in the present case allows for some couplings to be initially driven away from the fixed point that is ultimately reached. 

\begin{figure}
			\centering
			\begin{minipage}{0.8\linewidth}
				\raggedleft
				\includegraphics[clip, scale=.909]{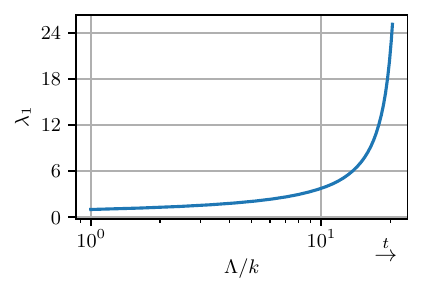}\\
				\includegraphics[clip, scale=.909]{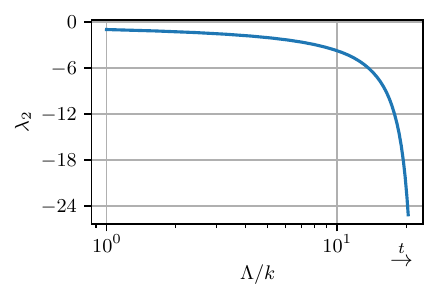}\\
				\includegraphics[clip, scale=.909]{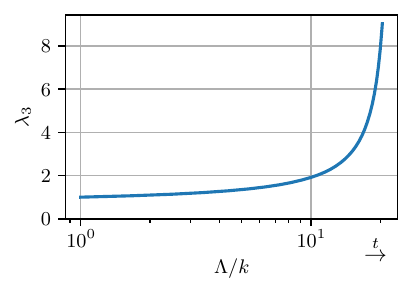}
			\end{minipage}
			\caption[Flow of the four-fermion couplings in purely fermionic dynamics.]{
					Flow of the four-fermion couplings in purely fermionic dynamics. 
					At ${ t = 0 }$ initial conditions ${ \lambda_i ( k = \Lambda) = (-1.0, -1.0, 1.0) }$ are chosen. 
			}
			\label{fig:fast_fermionic_flow_four_ferm_coupling}
		\end{figure}

\begin{figure}
			\centering
			\begin{minipage}{0.8\linewidth}
				\raggedleft
				\includegraphics[clip, scale=.909]{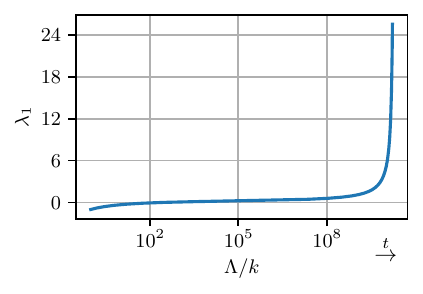}\\
				\includegraphics[clip, scale=.909]{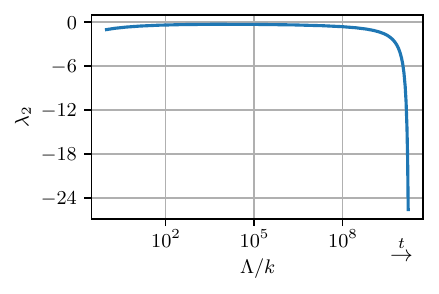}\\
				\includegraphics[clip, scale=.909]{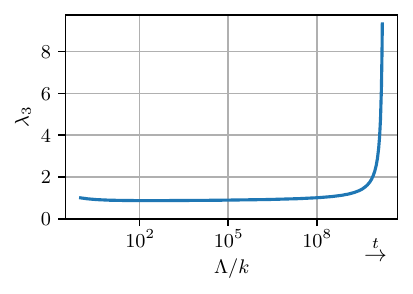}
			\end{minipage} 
			\caption[Flow of the four-fermion couplings in purely fermionic dynamics.]{
					Flow of the four-fermion couplings in purely fermionic dynamics. 
					At ${ t = 0 }$ initial conditions ${ \lambda_i ( k = \Lambda) = (-1.0, -1.0, 1.0) }$ are chosen. 
			}
			\label{fig:slow_fermionic_flow_four_ferm_coupling}
		\end{figure}

	The other scenario is that the couplings diverge at a certain scale. 
	They approach $ \pm \infty $. 
	This is also known from the \gls{gn} and related models. 
	Here, however, the mutual coupling of the equations makes it intricate to determine the precise parameter space for this or the fixed point scenario. 
	Two examples are shown in \cref{fig:fast_fermionic_flow_four_ferm_coupling,fig:slow_fermionic_flow_four_ferm_coupling}. 
	The example in \cref{fig:slow_fermionic_flow_four_ferm_coupling} illustrates once more that very large \gls{rg} times must sometimes be resolved.

	The interesting question is of course whether the gauge dynamics drive the couplings to values which either diverge or approach a fixed point due to the purely fermionic dynamics. 
	The observation is that the case of the divergence is relevant to \gls{twoqcd}. 
	It seems that the fixed point scenario requires specially balanced coupling values while the divergence is more generic in the Fierz-complete model as the divergence is close to the underlying properties of the \gls{gn}-type model. 

	The growing fermionic self-interactions ending up in a singularity have an important physical significance. 
	It signals the possible formation of condensates and mesonic bound states/resonances. 
	This can be best understood by means of a Hubbard--Stratonovich transformation \cite{Hubbard:1959ub,Stratonovich:1957}, where one treats the quark bilinears as auxiliary bosonic fields coupled to the quarks via a Yukawa interaction. 
	This method is well-known in condensed matter and high energy physics, see, \eg, \Reff\cite{Metzner:2011cw} for a review. 
	What is important is that the four-fermion couplings are inversely proportional to the mass of the bosonic field. 
	If it approaches its singularity, mesonic particles are produced with very little energy and, eventually, it signals the formation of a condensate. 
	The fermionic interactions can no longer be approximated as being local. 
	Instead, momentum dependence must be resolved for example by methods of partial bosonization, which is further outlined in the outlook, \cref{sec:outlook}.
	In other words, bound states form during the flow but their mass decreases for increasing four-fermion coupling making their propagation non-negligible.
	Hence, one might argue that the divergence of the four-fermion couplings already signals the limitations of the approximation and range of validity and suggests the path to a more sophisticated truncation.

	We recap this investigation with focus on the relevance of the dimension ${ d = 2 }$ and the Fierz completeness of our ansatz. 
	The purely fermionic dynamics lead to either a divergence of the four-fermion couplings or to a fixed point where most couplings become zero. 
	The former case seems to be the relevant scenario for \gls{twoqcd}. 
	In two dimensions, there is no critical absolute value to which the couplings must be driven to enter this scenario, which is different from ${3 + 1}$ dimensional \gls{qcd}.

\subsubsection{Infinite-\texorpdfstring{$ \Nc $}{Nc} limit }	

	Another interesting limit is the infinite-$ \Nc $ limit as already motivated and discussed for the minimal ansatz in \cref{sec:large_Nc_limit}. 
	In addition, many models are exactly solvable in this limit and different questions can be addressed analytically. 
	For instance, the 't Hooft solution \cite{tHooft:1974pnl} reveals the bound states of the theory while the \gls{gn} model can be solved to obtain even its phase diagram \cite{Thies:2003kk,Thies:2006ti}, see \Reffs\cite{Basar:2009fg,Boehmer:2009sw,Thies:2019ejd} for more examples. 

	There is no further rescaling of the couplings ${ \lambda_1, \dots, \lambda_3 }$ necessary because there is already a factor $ 1/\Nc $ included in the definition in \cref{eq:four-fermion-ansatz}. 
	On a diagrammatic level, the two top diagrams in \cref{fig:box_diagrams_four_int,fig:tri_diagrams_four_int} as well as the right one in \cref{fig:ferm_diagrams_four_int} are suppressed in this limit. 
	As we will see, the contributions from the remaining diagrams also simplify significantly. 
	For explicit expressions for the equations we refer to \Reff\cite[App.~I]{Oevermann:2024thesis}. 

	We find that the flow of the third channel vanishes completely in this limit. 
	This is presumably a regulator feature in combination with the spacetime dimension ${ d = 2 }$ as will be discussed in \cref{sec:regulator_artefacts}, \cf{} also \Reff\cite[Sec.~III.B]{Cresswell-Hogg:2025wda}.

	The major observation, however, is that the flow equations decouple such that the flow of $ \lambda_i $ does exclusively depend on the gauge coupling, the fermion mass and the coupling $ \lambda_i $ itself.
	We emphasize that this holds on all scales and it is a feature of the chosen basis of interactions. 
	Not only does this happen because the mentioned diagrams drop out, but also the remaining diagrams simplify.
	Similar observations have been made in \Reff\cite{Braun:2011pp} for the (${ 3 + 1 }$)-dimensional case.

	There is no analytical solution to the full system of flow equations because of the mass term. 
	The purely fermionic diagrams, however, simplify to the \gls{gn} equations which is investigated in detail in \cref{eq:Infinite_Nc_limit_of_four-fermion_flow_equations}.
	The \gls{gn} equations admit an exact solution, see \textit{e.\,g.}, \Reff\cite[Sec.~9]{Koenigstein:2023wso}.

	The conclusion is that it would be possible to work in a non Fierz-complete basis in this limit because the equations decouple.

\subsection{Regulator dependences}\label{sec:regulator_artefacts}

	This section summarizes a few consistency checks that have been carried out using a different regulator for the fermions, namely the Litim regulator \cite{Litim:2001up},
		\begin{align}\label{eq:Litim_regulator_insertion}
			\Delta S_k = \int_q \bpsi ( - q ) \,  \slashed{q} \, r_\mathrm{ f }( q ) \, \psi ( q ) \, , \vdistance
		\end{align}
	where the Litim regulator shape function is given by 
		\begin{align}
			r_\mathrm{ f }( q ) \equiv \bigg( \sqrt{ \frac{ k^2 }{ q^2 } } - 1 \bigg) \, \Theta \bigg( \frac{ k^2 }{ q^2 }- 1 \bigg) \, , \Vdistance
		\end{align}
	where $ \Theta $ is the Heaviside step function. 

	It has major deficiencies compared to the momentum-independent regulator \labelcref{eq:regulator_choice}, which is why it was not chosen in this work. 
	First of all, the momentum dependence destroys gauge invariance. 
	Only covariant derivatives should appear. 
	Moreover, it introduces crucial nonanalyticities at vanishing momentum and ruins spectral properties. 
	Unitarity is broken and the Osterwalder--Schrader reflection positivity is violated in the Euclidean domain. 
	On the other hand, the Litim regulator preserves axial symmetry, because it is proportional to $ \gamma^\mu $, which is a significant advantage and can be used for some consistency checks.

	Repeating the calculations with this regulator, we confirm that axial symmetry is indeed preserved: 
	setting the bare quark mass to zero, its flow vanishes and one indeed has ${ \partial_t ( \lambda_1 + \lambda_2 ) = 0 }$ from all types of diagrams if ${ \lambda_1 = -\lambda_2 }$ holds at a particular scale, \eg{} in the \gls{uv}. 

	It is possible to draw a few more conclusions. 
	The flow of the four-fermion couplings also decouples with this regulator in the infinite-$ \Nc $ limit suggesting that this is not a unique feature of the local regulator. 
	Moreover, the contributions from the box- and triangle-diagrams to the flow of the coupling $ \lambda_3 $ also vanish in this limit. 
	But there is a contribution from the purely fermionic diagrams in \cref{fig:ferm_diagrams_four_int}, which is a term proportional to $  ( m / k ) \, \lambda_3^2 $. 
	Therefore, one has to be careful about drawing strong conclusions from the simplification of the flow equation of $ \lambda_3 $, \cref{eq:flow_zero_gauge_coupling_three}.

	However, a detailed analysis of using other regulators is beyond the scope of this work and might be subject to future investigations. 
	Moreover, we highlight again the importance of property \labelcref{item:locality} that is obeyed by the local regulator and why we focus on it in this work. 

\subsection{Contribution to quark two-point vertex}\label{sec:contribution_to_fermion_two_pt_vertex}

	So far, it was neglected that the four-fermion couplings also drive the flow of the full fermion two-point vertex function.
	Equation \labelcref{eq:flow_ferm_2PT_vertex_diagrams} therefore obtains a new term which is of ``tadpole''-type,
		\begin{align}\label{eq:flow_ferm_2PT_vertex_diagrams_tadpole}
			\partial_t 
			\begin{gathered}
				\includegraphics[clip, trim = .5cm .5cm 3.2cm .5cm]{anc/flow_ferm_two_pt.pdf}
			\end{gathered}
			= \, &
			\begin{gathered}
				\includegraphics[clip, trim = 3.3cm 0cm .6cm .1cm]{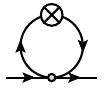}
			\end{gathered}
			+ \begin{gathered}
				\includegraphics[clip, trim = 0cm 0cm 2.7cm 0cm]{anc/ferm_two_pt_gluon_regulator.pdf}
			\end{gathered}\notag
			\\
			& +
			\begin{gathered}
				\includegraphics[clip, trim = 0cm 0cm 2.7cm 0cm]{anc/ferm_two_pt_fermion_regulator.pdf}
			\end{gathered} \, .
		\end{align}
	The first diagram on the \gls{rhs} is independent of the external momentum meaning that there is no contribution to the anomalous dimension. 
	This is because a projection onto the kinetic term consists of taking a derivative \gls{wrt}~the external momentum yielding zero. 
	As already mentioned in \cref{sec:consistency_check_quark_gluon_vertex}, a vanishing anomalous dimension $ \eta_\psi $ implies that the projection of the \gls{rhs}\ of the Wetterich equation onto the fermion-gluon vertex must give zero.
	However, the four-fermion vertex allows for an additional diagram, which was not present before, 
		\begin{align}\label{eq:fermion_gluon_three_pt_tadpole_fey_diagram}
			& \vcenter{ \hbox{  \includegraphics[clip, trim = 2.3cm .1cm 0cm .1cm]{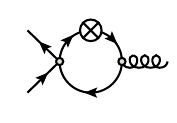} } } \, . \Vdistance
		\end{align}
	We verified explicitly that its projection gives no contribution. 
	However, \cref{eq:flow_ferm_2PT_vertex_diagrams_tadpole} gives a new contribution to the flow of the fermion mass. 
	Because the four-fermion couplings are dimensionless and we use a mass-like regulator, the diagram is logarithmically \gls{uv}-divergent. 
	Even though the couplings are initialized at zero, fluctuations always contribute on all scales with a mass-like regulator. 
	Formulae on the divergence and how the diagram can be regularized by an additional Pauli--Villars regularization are given in \Reff\cite[Sec.~5.5]{Oevermann:2024thesis}. 
	
	A solution to this problem might be to resolve the momentum dependence of the four-fermion couplings, which is expected to suppress the large-momentum contributions in the loop.
	Consequently, the problem certainly arises because of our (a bit too) simplistic truncation. 
	The diagrams in \cref{fig:box_diagrams_four_int,fig:tri_diagrams_four_int,fig:ferm_diagrams_four_int} of course carry a momentum dependence. 
	It is partly worked out in \Reff\cite[Sec.~G.3]{Oevermann:2024thesis} and we anyway have to resolve it to deal with the divergence of the four-fermion couplings. 
	We discuss these upcoming steps of our research program in the following outlook.

\section{Outlook: bosonization}\label{sec:outlook}

	The next step for improving the truncation in the matter sector is to resolve the momentum dependence of the four-fermion couplings. 
	It is the key to understanding the \gls{ir} physics of the model and to accessing the formation of (mesonic) bound states. 
	This can be pursued either on the level of the four-fermion couplings themselves, as done in \Reffs\cite{Fu:2022uow,Fu:2024ysj,Fu:2025hcm} in four dimensions with certain approximations, or by means of a ``flowing bosonization''. 
	This technique maps the nonlocalities of the four-fermion vertices to a local description in terms of a Yukawa coupling and meson exchanges \cite{Gies:2006wv}. 
	Diagrammatically speaking this means
		\begin{align}\label{eq:bosonisation_diagrams}
			& \begin{gathered}
				\includegraphics[clip, trim = 1.5cm 0cm .1cm 0cm]{anc/four_ferm_vertex.pdf}
			\end{gathered}
			\to
			\\
			& \begin{gathered}
				\includegraphics[clip, trim = 2.7cm 0cm .2cm 0cm]{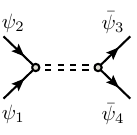}
			\end{gathered}
			+ 
			\begin{gathered}
				\includegraphics[clip, trim = 2.9cm 0cm .3cm 0cm]{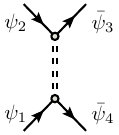}
			\end{gathered}
			+
			\begin{gathered}
				\includegraphics[clip, trim = 2.9cm 0cm .3cm 0cm]{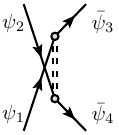}
			\end{gathered}
			+ \dots \, , \notag
		\end{align}
	where the dots stand for those parts of the momentum dependent vertex which cannot be written in terms of meson exchanges (dotted double lines) and a Yukawa coupling. 
	The Yukawa coupling will be of positive energy dimension and the ``tadpole''-type contribution to the fermion two-point vertex in \cref{eq:flow_ferm_2PT_vertex_diagrams_tadpole} will be replaced by self-energy diagrams as in \cref{eq:flow_ferm_2PT_vertex_diagrams}, but with meson instead of gluon lines. 
	This will lead to \gls{uv}-finiteness in all flow equations. 

	A challenge is that this formulation cannot be achieved by a one-off Hubbard--Stratonovich transformation, but it must be implemented continuously at all \gls{rg}-scales for which \Reffs\cite{Gies:2001nw,Pawlowski:2005xe,Floerchinger:2009uf,Ihssen:2024ihp} provide a recipe. 
	A great advantage of working in two dimensions compared to four is that there are much fewer possible bound states. 
	Hence, we aim to bosonize each channel in future work. 

	At the same time, we plan to improve the truncation in the gauge sector to remove the divergence of the gauge coupling to, at least, significantly lower scales compared to the ones where the four-fermion couplings diverge due to the purely fermionic dynamics.

\section{Summary and conclusion}\label{sec:summary}

	In summary, we have studied \gls{qcd} with massive quarks in the fundamental representation of the gauge group, in ${ d = 2 }$ spacetime dimensions. 
	Our starting point was the Euclidean functional integral and generating functionals for quarks and gauge fields, as well as ghosts after gauge fixing. 
	The latter was done using the background field variant of covariant gauges (and we concentrated on Feynman--'t Hooft gauge for the numerical part of our calculations).

	In a subsequent step we introduced a mass-like \gls{ir} regularization for all fluctuating fields. 
	When the corresponding regulator scale is very large, quantum fluctuations are suppressed, while they get included successively when the regularization scale is lowered. 
	We carefully examined how the presence of the \gls{ir} regularization modifies generating functionals, and in particular identities related to gauge invariance they obey: Ward--Takahashi, Slavnov--Taylor, and Nielsen. 
	Because \gls{uv} divergences are very mild in ${ d = 2 }$ dimensions, we can work with a momentum-independent regulator that preserves the Ward-Takahashi identity in an unmodified way, and the Nielsen identity also has no explicit regulator dependence.
	\Gls{brst} symmetry is broken, and the Slavnov--Taylor identity is accordingly modified, however.

	As a super-renormalizable theory, \gls{twoqcd} is well controlled in the \gls{uv} regime. 
	The gauge coupling, which has dimensions of energy, goes to a fixed value and a \gls{uv} regularization can be removed. 
	The \gls{ir}, on the other side, is more problematic. 
	To get started, we made a relatively simple ansatz, or truncation, for the flowing quantum effective action with only two flowing parameters, the gauge coupling and the fermion mass. 
	The wavefunction renormalization factor for quarks turned out not to flow within the truncation and gauge fixing we have chosen.

	Being close to the perturbative framework, our truncation allowed to reproduce the perturbative one-loop result for the beta functions, but also already contains non-perturbative elements beyond that. 
	In the \gls{ir} regime, we find that the gauge coupling and, as a consequence, also the fermion mass diverge. 
	One should be careful in drawing too many conclusions from this finding because the flow equations for \gls{qcd} in ${ d = 4 }$ dimensions would likely behave similarly for a truncation as simple as ours. 
	Further investigations are needed before definite conclusions can be drawn concerning the \gls{ir} sector.

	One obvious question concerns the dependence on the gauge fixing. 
	Besides Feynman--'t Hooft gauge it would be useful to study other covariant gauges in future work. 
	We are also planing to explore light-cone gauge further, especially to connect more directly to the pioneering work of 't Hooft.

	Within our present approximation, it was also interesting to vary the gauge group through the number of color charges $ \Nc $, and the number of quark flavors $ \Nf $. 
	When lowering the ratio of the number of gauge bosons to fermions below a certain threshold, we found that there is a window in the \gls{rg} flow, where the gauge coupling first decreases, before it ultimately also diverges towards the \gls{ir} for massive quarks. 
	Only for an infinite number of fermionic flavors it is possible to fully integrate the flow to the \gls{ir} and to fully remove the regulator in our current truncation, and to arrrive in a weakly coupled regime.

	Going beyond gauge field dynamics, we also investigated local fermion-fermion interaction terms of various kinds. 
	These are generated by the \gls{rg} flow from gluon and quark field fluctuations. 
	This basic mechanism is well known from \gls{qcd} in ${ d = 4 }$ dimensions and still works in ${ d = 2 }$ despite the fact that there are no propagating gluons. 

	Via two different approaches, (1.) expansion in the full tensor structure and (2.) projection and algebraically solving for the \gls{rg} flows of the single couplings, we explained and worked out that there are only three algebraically independent local fermion interaction terms in the case of one flavor, ${ \Nf = 1 }$. 
	Chiral symmetry for massless quarks would reduce this further to two. 
	We also included the feedback of these local fermion interaction couplings to the flow, once they have been generated. 
	Also within this extended truncation there is still the \gls{ir} divergence stemming from the gauge coupling at which the flow breaks down.

	We observed that the fermion-gluon dynamics drive the four-fermion couplings to large values and ultimately to a divergence. 
	This divergence is not necessarily due to the divergence of the gauge coupling. It can also occur in purely fermionic theories with frozen or vanishing gauge coupling once the fermion-fermion couplings are nonzero. 
	Divergent local four-fermion couplings should be interpreted as a first signal for the formation of bound states and the relevance of the respective interaction channels -- at least at intermediate \gls{rg} scales. 
	We remark that, even though chiral symmetry is broken by the fermionic part of the Callan--Symanzik regulator, the \gls{rg} flow of the four-fermion couplings produces the correct signs of the couplings that would be required for full chiral symmetry. 

	In upcoming work, we plan to better resolve the frequency and momentum dependence of quark interaction vertices and especially to resolve the branch cuts and poles related to resonances and bound states. 
	The precise determination of the spectrum of bound states and resonances through the \gls{rg} flow is the most important motivation and future goal for us.

\section*{Acknowledgements}

	We thank Holger Gies for useful discussions and Markus Huber for support with the \dofun{} software-package. 
	S.\,F. thanks Moaathe Belhaj Ahmed for collaboration on related work and E.\,O. acknowledges support from the Stiftung der Deutschen Wirtschaft (sdw).

\section*{Data availability}

	The data that support the findings of this article are openly available \cite{oevermann2024functionalrenormalizationqcd1}.

\appendix

\section{BRST symmetry}\label{app:BRST_symmetry}

	Here, we formulate the \gls{brst}-symmetry \cite{Becchi:1975nq,Tyutin:1975qk} within the background-field method similar to \Reff\cite[Sec.~3.2.1]{Gkiatas:2023rtn}, see also \Reff\cite[Sec.~16.4]{Peskin:1995ev} for the case without a background-field. 
	To do so, we introduce non-dynamical, auxiliary fields $ B^z $, $ { z \in \{ 1, \ldots, \Nc^2 - 1 \} } $, oftentimes denoted as Nakanishi-Lautrup fields \cite{Nakanishi:1966zz,Lautrup:1967zz},
		\begin{align}\label{eq:action_with_B_field}
			& S [ \bpsi, \psi, a, \bA, \bar{ c }, c ] \vdistance
			\\
			= \, &  \int_x \Big[ 
				\bpsi \, ( \gamma^\mu \, D_\mu + m ) \, \psi 
				+ \frac{ 1 }{ 2 \coupling^2 } \, \tr ( F_{\mu\nu} \, F^{\mu\nu} ) \vdistance  \notag
			\\
			& - \bar{ c } \, D^\mu [ \bA ] \, D_\mu [ a + \bA ] \, c  + \frac{ \xi }{ 2 } \, B^2
				+ \frac{ 1 }{ \coupling } \, B \, D^\mu [ \bA ] \, a_\mu
			\Big]
			 \, , \vdistance \notag
		\end{align}
	over which we also integrate in the path integral. 
	Hence, \cref{eq:gauge_fixed_microscopic_action,eq:action_with_B_field} are equivalent because $ B $ can simply be integrated out. 
	A global \gls{brst} transformation with infinitesimal parameter $ \epsilon $ is given by 
		\begin{alignat}{3}
			& \bar{A}_\mu^z && \mapsto \bar{A}^{\prime z}_\mu && = \bar{A}_\mu^z + 0 \, , \vdistance
			\\
			& a_\mu^z && \mapsto a^{\prime z}_\mu && = a_\mu^z + \epsilon \, D_\mu [ a + \bar{A} ]\indices{^z_w} \, c^w \, , \vdistance
			\\
			& \psi && \mapsto \psi' && = \psi + \ii \, \epsilon \, c^z \, T_z \, \psi \, , \vdistance 
			\\
			& \bpsi && \mapsto \bpsi' && = \bpsi - \ii \, \epsilon \, \bpsi \, c^z \, T_z \, , \vdistance
			\\
			& c^z && \mapsto c^{\prime z} && = c^z - \frac{ 1 }{ 2 } \, \epsilon \, f\indices{^z_{ u v }} \, c^u \, c^v \, , \vdistance 
			\\
			& \bc^z && \mapsto \bc^{\prime z} && = \bc^z + \epsilon \, B^z \, , \vdistance
			\\
			& B^z && \mapsto B^{\prime z} && = B^z + 0 \, . \vdistance
		\end{alignat}
	The fluctuation-field $ a $ carries the \gls{brst} transformation in these formulas. 
	It is not the only possible way of distributing the \gls{brst} transformation over the fields \cite[Sec.~3.2.1]{Gkiatas:2023rtn}. 
	The \gls{brst}-invariance is of importance because it must also be present in the quantum effective action.

\section{Ward identity for CS-type regulators}\label{app:Modified_Ward--Takahashi_identity}

	In presence of an arbitrary regulator piece, the Ward--Takahashi identity \labelcref{eq:WT_identity} becomes the \textit{modified Ward--Takahashi identity} for $ \Gk $ \cite[Eq.~(74)]{Gies:2006wv}, 
		\begin{align}\label{eq:modified_WT_identity}
			\mathcal{ G }^z ( \Gk + \Delta S_k ) = \langle \mathcal{ G }^z ( S_{ \mathrm{gf} } + S_{ \mathrm{gh} } + \Delta S_k ) \rangle_{ J [ \Phi ] } \, .
		\end{align}
	Recall that $ \Gk $ is not yet evaluated at $ {\bA = A} $ and $ {a = 0} $. 
	The regulator based modification, ${ \langle \mathcal{ G }^z \Delta S_k \rangle - \mathcal{ G }^z \Delta S_k }$, vanishes for the local regulator \labelcref{eq:regulator_choice}. 
	We partly follow \Reff\cite[Sec.~3.4]{Gies:2006wv} to prove this. 
	The quark and ghost parts are individually invariant under gauge transformations. 
	Focusing on the gauge field only and using \cref{eq:generator_gauge_transformation}, we calculate
		\begin{align}
			& D_\mu [ a + \bA ]\indices{^z_w} \, \frac{ \updelta }{ \updelta a_{ \mu w } ( x ) } 
				\int_y \frac{ k^2 }{\coupling^2} \, \tr [ a_\mu ( y ) \, a^\mu ( y ) ] \Vdistance
			\\
			= \, & \frac{k^2}{\coupling^2} \, \Big(
				\updelta_w^z \, \partial_\mu + f\indices{^z_{uw}} \, [ a_\mu^u ( x ) + \bA_\mu^u ( x ) ]
			\Big) \, 
			a^{ \mu w } ( x ) \Vdistance \notag
			\\
			= \, & \frac{k^2}{\coupling^2} \, [ \partial_\mu a^{ \mu z } ( x ) + f\indices{^z_{ u w }} \, \bA_\mu^u ( x ) \, a^{ \mu w } ( x ) ] \, . \Vdistance \notag
		\end{align}
	The final expression is linear in the fluctuation-field which implies that both regulator terms in \cref{eq:modified_WT_identity} automatically cancel. 
	The key property of the Callan--Symanzik-type regulator that leads to this result is that it is diagonal in color space.

\section{Derivation of the modified Nielsen identity}\label{app:Modified_Nielsen_identity}

	We derive \cref{eq:modified_Nielsen_identity_CZ} and a more general result for arbitrary regulator pieces using the field space vectors
		\begin{align}
			\Phi_\mathbf{ a } = \, & ( \psi, \bpsi, a, c, \bar{ c } )_\mathbf{ a } \, , \vdistance
			\\
			\bar{ \Phi }_\mathbf{ a } = \, & ( \psi, \bpsi, a + \bA, c, \bar{ c } )_\mathbf{ a } \, , \vdistance
			\\
			J^\mathbf{ a } = \, & ( \bar{ \eta }, -\eta, J, \bar{ \omega }, -\omega )^\mathbf{ a } \, . \vdistance
		\end{align}
	As an auxiliary calculation, we need the functional derivative of the Schwinger functional \gls{wrt}\ the background field at fixed source,
		\begin{align}\label{eq:derivative_Schwinger_functional_wrt_background_field}
			& \frac{ \updelta W_k  }{ \updelta \bA } \Big\vert_J =	\Vdistance
			\\
			= \, & \frac{ 1 }{ Z_k } \, \int \mathcal{ D } \Phi \, \frac{ \updelta }{ \updelta \bA } \, \ee^{ - ( S_{\text{QCD}_2} + S_{ \mathrm{gf} } + S_{ \mathrm{gh} } ) + J^{ \mathbf{ a } } \, \bar{ \Phi }_{ \mathbf{ a } } - \Delta S_k }	\Vdistance	\notag
			\\
			= \, & \frac{ 1 }{ Z_k } \, \int \mathcal{ D } \Phi \, \ee^{ - ( S_{ \mathrm{gf} } + S_{ \mathrm{gh} } + \Delta S_k ) } \, \frac{ \updelta }{ \updelta \bA } \, \Big( \ee^{ - S_{\text{QCD}_2} + J^{ \mathbf{ a } } \, \bar{ \Phi }_{ \mathbf{ a } } } \Big)	\Vdistance	\notag
			\\
			& - \Big\langle \frac{ \updelta ( S_{ \mathrm{gf} } + S_{ \mathrm{gh} } ) }{ \updelta \bA } \Big\rangle	\Vdistance	\notag
			\\
			& - \frac{ 1 }{ Z_k } \, \int \mathcal{ D } \Phi \, \frac{ \updelta \Delta S_k }{ \updelta \bA } \, \ee^{ - ( S_{\text{QCD}_2} + S_{ \mathrm{gf} } + S_{ \mathrm{gh} } ) + J^{ \mathbf{ a } } \, \bar{ \Phi }_{ \mathbf{ a } } - \Delta S_k  }	\Vdistance	\notag
			\\
			= \, & \frac{ 1 }{ Z_k } \, \int \mathcal{ D } \Phi \, \ee^{ - ( S_{ \mathrm{gf} } + S_{ \mathrm{gh} } + \Delta S_k ) } \, \frac{ \updelta }{ \updelta a } \, \Big( \ee^{ - S_{\text{QCD}_2} + J^{ \mathbf{ a } } \, \bar{ \Phi }_{ \mathbf{ a } } } \Big)	\Vdistance	\notag
			\\
			& - \Big\langle \frac{ \updelta ( S_{ \mathrm{gf} } + S_{ \mathrm{gh} } ) }{ \updelta \bA } \Big\rangle \, - \frac{ 1 }{ 2 } \, \Big[ \mathrm{STr} \Big( \frac{ \updelta R_k }{ \updelta \bA} \, G_k \Big) + \Phi \, \frac{ \updelta R_k }{ \updelta \bA } \, \Phi  \Big]	\Vdistance	\notag
			\\
			= \, & \frac{ 1 }{ Z_k } \, \int \mathcal{ D } \Phi \, \frac{ \updelta }{ \updelta a } \,  \Big( \ee^{ - ( S_{\text{QCD}_2} + S_{ \mathrm{gf} } + S_{ \mathrm{gh} } + \Delta S_k ) + J^{ \mathbf{ a } } \, \bar{ \Phi }_{ \mathbf{ a } } } \Big)	\Vdistance	\notag
			\\
			& + \Big\langle \frac{ \updelta ( S_{ \mathrm{gf} } + S_{ \mathrm{gh} } ) }{ \updelta a } \Big\rangle + \frac{ \updelta \Delta S_k }{ \updelta a } - \Big\langle \frac{ \updelta ( S_{ \mathrm{gf} } + S_{ \mathrm{gh} } ) }{ \updelta \bA } \Big\rangle	\Vdistance	\notag
			\\
			& - \frac{ 1 }{ 2 } \, \mathrm{STr} \Big( \frac{ \updelta R_k }{ \updelta \bA} \, G_k \Big) - \frac{ 1 }{ 2 } \, \Phi \, \frac{ \updelta R_k }{ \updelta \bA } \, \Phi	\Vdistance	\notag
			\\
			= \, & \Big\langle \frac{ \updelta ( S_{ \mathrm{gf} } + S_{ \mathrm{gh} } ) }{ \updelta a } - \frac{ \updelta ( S_{ \mathrm{gf} } + S_{ \mathrm{gh} } ) }{ \updelta \bA } \Big\rangle \, + \frac{ \updelta \Delta S_k }{ \updelta a }	\Vdistance	\notag
			\\
			& - \frac{ 1 }{ 2 } \, \mathrm{STr} \Big( \frac{ \updelta R_k }{ \updelta \bA} \, G_k \Big) - \frac{ \updelta \Delta S_k }{ \updelta \bA} \, .	\Vdistance	\notag
		\end{align}
	Using this result, we obtain
		\begin{align}
			& \frac{ \updelta \Gk }{ \updelta \bA } \Vdistance
			\\
			= \, & \int_x \Big( \cancel{ \frac{ \updelta J^\mathbf{ a } }{ \updelta \bA } \, \bar{ \Phi }_\mathrm{ a } } + J^\mathbf{ a } \, \frac{ \updelta \bar{ \Phi }_\mathbf{ a } }{ \updelta \bA } \Big) - \cancel{ \frac{ \updelta J^\mathbf{ a } }{ \updelta \bA } \, \frac{ \updelta W_k }{ \updelta J^\mathbf{ a } } } \Vdistance \notag 
			\\
			& - \frac{ \updelta W_k }{ \updelta \bA } \Big\vert_J -  \frac{ \updelta \Delta S_k }{ \updelta \bA } \stackrel{\labelcref{eq:derivative_Schwinger_functional_wrt_background_field}}{=}	\Vdistance	\notag
			\\
			= \, & \frac{ \updelta \Gk }{ \updelta a } + \frac{ \updelta \Delta S_k }{ \updelta a } - \Big\langle \frac{ \updelta ( S_{ \mathrm{gf} } + S_{ \mathrm{gh} } ) }{ \updelta a } - \frac{ \updelta ( S_{ \mathrm{gf} } + S_{ \mathrm{gh} } ) }{ \updelta \bA } \Big\rangle	\Vdistance	\notag
			\\
			& - \frac{ \updelta \Delta S_k }{ \updelta a } + \mathrm{STr} \Big( \frac{ \updelta R_k }{ \updelta \bA} \, G_k \Big) \, .	\Vdistance \notag
		\end{align}
	Re-arranging the terms yields the result
		\begin{align}\label{eq:modified_Nielsen_identity}
			& \frac{ \updelta \Gk }{ \updelta \bA } - \frac{ \updelta \Gk }{ \updelta a } + \bigg \langle \frac{ \updelta ( S_{ \mathrm{gf} } + S_{ \mathrm{gh} } ) }{ \updelta a } - \frac{ \updelta ( S_{ \mathrm{gf} } + S_{ \mathrm{gh} } ) }{ \updelta \bA } \bigg \rangle \Vdistance \notag
			\\
			= \, & \frac{ 1 }{ 2 } \, \mathrm{STr} \bigg( \frac{ \updelta R_k }{ \updelta \bA} \, G_k \bigg) \, . \Vdistance 
		\end{align}
	On-shell evaluation with sources 
		\begin{align}\label{eq:source_on_shell}
			J^\mathbf{ a } = \frac{ \updelta \Delta S_k [ \Phi ] }{ \updelta \Phi_\mathbf{ a } }
		\end{align}
	leads to ${ \updelta \Gk / \updelta a = 0 }$ and a vanishing of the expectation value on the \gls{lhs} of \cref{eq:modified_Nielsen_identity},
		\begin{align}
			& \bigg \langle \frac{ \updelta ( S_{ \mathrm{gf} } + S_{ \mathrm{gh} } ) }{ \updelta a } - \frac{ \updelta ( S_{ \mathrm{gf} } + S_{ \mathrm{gh} } ) }{ \updelta \bA } \bigg \rangle_{ J = \frac{ \updelta \Delta S_k }{ \updelta \Phi } } \Vdistance
			\\
			= \, & \frac{ 1 }{ Z_k } \, \int \mathcal{ D } \Phi \, 
				\Big(
					\frac{ \updelta }{ \updelta \bA } - \frac{ \updelta }{ \updelta a }
				\Big) \, \ee^{ - ( S_{\text{QCD}_2} + S_{ \mathrm{gf} } + S_{ \mathrm{gh} } ) }	\Vdistance	\notag
			\\
			= \, & \frac{ 1 }{ Z_k } \, \frac{ \updelta }{ \updelta \bA } \int \mathcal{ D } \Phi \, \ee^{ - ( S_{\text{QCD}_2} + S_{ \mathrm{gf} } + S_{ \mathrm{gh} } ) } = 0 \, .	\Vdistance	\notag
		\end{align}
	The last line is zero for the reason which is referred to as ``the Faddeev--Popov trick can be undone'' in \Reff\cite[Sec.~5.2.1]{Dupuis:2020fhh}. 
	The on-shell modified Nielsen identity therefore reads
		\begin{align}\label{eq:on_shell_modified_Nielsen_identity}
			& \frac{ \updelta \Gk }{ \updelta \bA } = \frac{ 1 }{ 2 } \, \mathrm{STr} \bigg( \frac{ \updelta R_k }{ \updelta \bA} \, G_k \bigg) \, . \Vdistance 
		\end{align}
	The \gls{rhs} is zero for vanishing cutoff, \cf{} \cref{eq:Nielsen_identity}, or the local regulator, \cf{} \cref{eq:modified_Nielsen_identity_CZ}.

\section{Rescaling of the gauge field}\label{app:Rescaling of the gauge field}

	There are two ways of introducing renormalized quantities in the gauge sector.
	For the first one the starting point is the bare action
		\begin{align}
			S = \frac{1}{2} \int \tr( F_{\text{bare}}^{\mu \nu} \, F_{\text{bare,} \mu \nu} )
		\end{align}
	where the field strength is built from the covariant derivative
		\begin{align}
			D_\mu = \partial_\mu - \ii \, g_{\text{bare}} \, A_{\text{bare,} \mu} \, .
		\end{align}
	Next, introduce a scale-dependent wave function renormalization
		\begin{align}
			S = \frac{Z_k}{2} \, \int \tr( F_{\text{bare}}^{\mu \nu} \, F_{\text{bare,} \mu \nu} ) \, .
		\end{align}
	Absorb it into the field
		\begin{align}
			\bar{A}_\mu = \, & \sqrt{ Z_k } \, A_{\text{bare,} \mu} \, , \vdistance
			\\
			S = \, & \frac{1}{2} \int \tr( \bar{F}^{\mu \nu} \, \bar{F}_{\mu \nu} ) \, , \vdistance
			\\
			D_\mu = \, & \partial_\mu - \ii \, \frac{ g_{\text{bare}} }{ \sqrt{ Z_k } } \, \bar{A}_{\mu} \, . \vdistance
		\end{align}
	Redefine the coupling to $ g_k $ such that
		\begin{align}
			D_\mu = \, & \partial_\mu - \ii g_k \, \bar{A}_{\mu} \, , \vdistance
			\\
			0 = \, & \partial_k ( g_k \, \sqrt{ Z_k } ) \, , \label{eq:eta_condition}
		\end{align}
	and absorb it into the field
		\begin{align}
			S = \, & \frac{1}{2 g_k^2} \, \int \tr( F^{\mu \nu} \, F_{\mu \nu} ) \, . \vdistance
		\end{align}
	Putting all steps together, one has not rescaled with a scale-dependent quantity, but just with $ g_{\text{bare}} $. 
	The two terms in the flow equation cancel due to \cref{eq:eta_condition}. 
	Because of that identity there is in fact just one independent renormalization parameter. 
	\newline
	The starting point of the second approach is again the bare action
		\begin{align}
			S = \frac{1}{2} \, \int \tr( F_{\text{bare}}^{\mu \nu} \, F_{\text{bare,} \mu \nu} ) \, .
		\end{align}
	One immediately absorbs the coupling in the gauge fields to remove the coupling from the gauge-covariant derivative,
		\begin{align}
			S = \, & \frac{1}{2 g_{\text{bare}}^2} \, \int \tr( F^{\mu \nu} \, F_{\mu \nu} ) \, , \vdistance
			\\
			D_\mu = \, & \partial_\mu - \ii A_{\mu} \, . \vdistance
		\end{align}
	Next, we allow the gauge coupling to be scale-dependent
		\begin{align}
			S = \, & \frac{1}{2 g_k^2} \, \int \tr( F^{\mu \nu} \, F_{\mu \nu} ) \, . \vdistance
		\end{align}
	Here, however, there is no additional need to introduce a $ Z_k $.
	Still, one has to introduce this term for the matter sector.
	In any case, the the two approaches are equivalent and both keep the structure of the gauge-covariant derivative intact.

\section{Infinite-\texorpdfstring{$ \Nc $}{Nc} limit of four-fermion flow equations}\label{eq:Infinite_Nc_limit_of_four-fermion_flow_equations}

    We investigate the purely fermionic dynamics in more detail. 
    In particular, we explain why the corresponding flow equations for the four-fermion couplings decouple in the infinite-$ \Nc $ limit for the specific choice of the Fierz-complete basis. 

    The first step towards this goal does not make use of these specifications yet. 
    We re-derive the flow equations for the four-fermion couplings in an arbitrary number of spacetime dimensions for a general basis $ \{ \mathcal{ O }_i \} $ of matrices such that $ \{ ( \bpsi \, \mathcal{ O }_i \, \psi )^2 \} $ is Fierz-complete.
    Therefore, we utilize the background-field method applied to the fermionic field: 
    we split it into a background-field $ \psi $, $ \bpsi $ and a fluctuation field $ \psi' $, $ \bpsi' $ and extract from the four-fermion interaction the term bilinear in the fluctuation fields,
        \begin{align}
            &   \big[ 
                    ( \bpsi + \bpsi' ) \, \mathcal{ O }_i \, ( \psi + \psi' )
                \big]^2 \vdistance 
            \\
            = \, &  \big[
                    \bpsi \, \mathcal{ O }_i \, \psi + \bpsi \, \mathcal{ O }_i \, \psi' + \bpsi' \, \mathcal{ O }_i \, \psi + \bpsi' \, \mathcal{ O }_i \, \psi'
                \big]^2 \vdistance \notag
            \\
            = \, &  2 \, ( \bpsi \, \mathcal{ O }_i \, \psi ) \, ( \bpsi' \, \mathcal{ O }_i \, \psi' ) + 2 \, ( \bpsi \, \mathcal{ O }_i \, \psi' ) \, ( \bpsi' \, \mathcal{ O }_i \, \psi ) + \dots \vdistance \notag
        \end{align}
    We proceed with the Wetterich equation in its $ \log $-form and expand its \gls{rhs} up to quartic order in the fermionic field to project on the flow of the four-fermion couplings. 
    Denoting the free-part of ${ ( \Gk^{ ( 2 ) } + \Delta S_k ) }$ by $ P $, we obtain (summation over $ i $, $ j $ is implied)
        \begin{align}
            & ( \partial_t \lambda_i + 2 \, \eta_\psi \, \lambda_i ) \, ( \bpsi \, \mathcal{ O }_i \, \psi ) \, ( \bpsi \, \mathcal{ O }_i \, \psi ) \vdistance
            \\
            = \, & \mathrm{STr} \Big[ \tilde{ \partial }_t \Big(
                \big[ \lambda_ i \, ( \bpsi \, \mathcal{ O }_i \, \psi ) \, \mathcal{ O }_i + \lambda_ i \, \mathcal{ O }_i \, \psi \, \bpsi \, \mathcal{ O }_i \big] \, P^{-1} \vdistance \notag
            \\
            & \times \big[ \lambda_j \, ( \bpsi \, \mathcal{ O }_j \, \psi ) \, \mathcal{ O }_j + \lambda_ j \, \mathcal{ O }_j \, \psi \, \bpsi \, \mathcal{ O }_j \big] \, P^{-1} \, 
                \Big) \Big] \vdistance \notag
            \\
            = \, & \lambda_i \, \lambda_j \, \Big[ 
                ( \bpsi \, \mathcal{ O }_i \, \psi ) \, ( \bpsi \, \mathcal{ O }_j \, \psi ) \, \mathrm{STr} \big( 
                    \tilde{ \partial }_t [ P^{-1} \, \mathcal{ O }_i \, P^{-1} \, \mathcal{ O }_j ]    
                \big)
                \vdistance \notag
            \\
            & + 2 \, ( \bpsi \, \mathcal{ O }_i \, \psi ) \, \mathrm{STr} \big( 
                    \tilde{ \partial }_t [ P^{-1} \, \mathcal{ O }_i \, P^{-1} \, ( \mathcal{ O }_j \, \psi \, \bpsi \, \mathcal{ O }_j ) ]    
                    \big)
                \vdistance \notag
            \\
            & + \mathrm{STr} \big( 
                \tilde{ \partial }_t [ P^{-1} \, ( \mathcal{ O }_i \, \psi \, \bpsi \, \mathcal{ O }_i ) \, P^{-1} \, ( \mathcal{ O }_j \, \psi \, \bpsi \, \mathcal{ O }_j ) ]    
                \big)
            \Big] \, . \vdistance \notag
        \end{align}
    The first term is already in the desired form of \gls{lhs} such that one can read off the flow equations by comparing both sides. 
    The other terms are not of this from yet. 
    In general, one would have to expand the matrix structure connecting $ \bpsi $ with $ \psi $ in terms of the basis $ \{ \mathcal{ O }_i \} $ in these last two lines to proceed with a comparison with the \gls{lhs}. 

    In the infinite-$ \Nc $ limit, however, only the first term is non-vanishing because it is the only one that generates a factor of $ \Nc $ in the trace. 
    Thus, what remains to be shown for the statement about the decoupling of the flow equations is that
        \begin{align}
            & \mathrm{STr} \big( 
                    \tilde{ \partial }_t [ P^{-1} \, \mathcal{ O }_i \, P^{-1} \, \mathcal{ O }_j ]    
                \big) \propto \updelta_{ i j } \, . \vdistance
        \end{align}
    For the choice of $ \mathcal{ O }_i \in \{ \openone, \gamma^\mu, \gammachiral \} $, this is indeed the case. 
    We split the free propagator into two parts,
        \begin{align}
            P^{ -1 } = a_k \, \openone + b_k \, p_{ \mu } \, \gamma^\mu \, , \vdistance
        \end{align}
    where the regulator derivative only acts on the scale-dependent coefficients $ a_k $ and $ b_k $. 
    We now separate the trace over Dirac space from the momentum integration. 
    We immediately see that
        \begin{align}
            \mathrm{tr} \big( 
                    \openone \, \mathcal{ O }_i \, \openone \, \mathcal{ O }_j
                \big) \propto \updelta_{ i j } \, . \vdistance
        \end{align}
    Moreover, linear parts in momentum vanish by integration if we evaluate at zero external momenta. 
    Similarly, integration simplifies $ p_{ \mu } \, p_{ \nu } \rightarrow \updelta_{ \mu \nu } \, p^2 / d $ such that we only need to prove $ \mathrm{tr} \big( 
                    \gamma_{ \mu } \, \mathcal{ O }_i \, \gamma^{ \mu } \, \mathcal{ O }_j
                \big) \propto \updelta_{ i j } $.
    \begin{itemize}
        \item For $ \mathcal{ O }_i = \openone $, this is clear using $ \gamma_{ \mu } \, \gamma^{ \mu } = d $. 
        \item For $ \mathcal{ O }_i = \gammachiral $, it follows from the first point and that the trace over an odd number of gamma-matrices vanishes. 
        \item For $ \mathcal{ O }_i = \gamma^{ \alpha } $, we can even infer that 
            \begin{align}
                \mathrm{tr} \big( 
                        \gamma_{ \mu } \, \gamma^{ \alpha } \, \gamma^{ \mu } \, \gamma^{ \beta }
                    \big) \propto (2 - d) \, \updelta^{ \alpha \beta } = 0 \, , \vdistance
            \end{align}
            which explains why there is no contribution to the flow of $ \lambda_3 $ in the infinite-$ \Nc $ limit.
        \end{itemize}

\FloatBarrier

%\nocite{*}
% produces the bibliography via BibTeX -- separated bibfiles for working versions
\bibliography{new_bib.bib,bib/general,bib/gn,bib/inhomo,bib/instanton,bib/lattice,bib/math,bib/numerics,bib/qcd,bib/rg,bib/software,bib/symmetries,bib/thermal_qft,bib/thies,bib/virasoro_algebra,bib/zero-dim-qft}

\end{document}